\documentclass[fleqn,usenatbib]{mnras}
\usepackage{amsmath}
\usepackage{mathptmx}
\usepackage{txfonts}

\usepackage[T1]{fontenc}
\usepackage{ae,aecompl}

\usepackage{graphicx}	% Including figure files
\usepackage{amssymb}	% Extra maths symbols
\usepackage{booktabs}
\usepackage{wasysym}
\usepackage{color}
\let\oldhref\href
\renewcommand{\href}[2]{\oldhref{#1}{\hbox{#2}}}

\definecolor{color1}{RGB}{0, 51, 204}
\definecolor{color2}{RGB}{204, 51, 0}
\definecolor{color3}{RGB}{180, 180, 0}
\definecolor{color4}{RGB}{0, 153, 0}
\definecolor{color5}{RGB}{153, 0, 153}

\newcommand{\hMpc}{{\ifmmode{h^{-1}{\rm Mpc}}\else{$h^{-1}$Mpc}\fi}}
\newcommand{\Mpc}{{\ifmmode{{\rm Mpc}}\else{Mpc}\fi}}
\newcommand{\hkpc}{{\ifmmode{h^{-1}{\rm kpc}}\else{$h^{-1}$kpc}\fi}}
\newcommand{\kpc}{{\ifmmode{ {\rm kpc} }\else{{\rm kpc}}\fi}}
\newcommand{\kms}{{\ifmmode{ {\rm km\,s^{-1}} }\else{ ${\rm km\,s^{-1}}$ }\fi}}
\newcommand{\hMsun}{{\ifmmode{h^{-1}{\rm {M_{\astrosun}}}}\else{$h^{-1}{\rm{M_{\astrosun}}}$}\fi}}
\newcommand{\Msun}{{\ifmmode{{\rm M}_{\astrosun}}\else{${\rm M}_{\astrosun}$}\fi}}
\newcommand{\Mhalo}{{\ifmmode{M_{\rm halo}}\else{$M_{\rm halo}$}\fi}}
\newcommand{\Rvir}{{\ifmmode{R_{\rm vir}}\else{$R_{\rm vir}$}\fi}}
\newcommand{\Mvir}{{\ifmmode{M_{\rm vir}}\else{$M_{\rm vir}$}\fi}}
\newcommand{\Mstar}{{\ifmmode{M_{\rm star}}\else{$M_{\rm star}$}\fi}}
\newcommand{\Vrot}{{\ifmmode{V_{\rm rot}}\else{$V_{\rm rot}$}\fi}}
\newcommand{\ltsima}{$\; \buildrel < \over \sim \;$}
\newcommand{\gtsima}{$\; \buildrel > \over \sim \;$}
\newcommand{\lsim}{\lower.5ex\hbox{\ltsima}}
\newcommand{\gsim}{\lower.5ex\hbox{\gtsima}}

\def\lesssim{\mathrel{\hbox{\rlap{\hbox{\lower4pt\hbox{$\sim$}}}\hbox{$<$}}}}
\def\gtrsim{\mathrel{\hbox{\rlap{\hbox{\lower4pt\hbox{$\sim$}}}\hbox{$>$}}}}

\newcommand{\beq}{\begin{equation}}
\newcommand{\eeq}{\end{equation}}
\def\beqa{\begin{eqnarray}}
\def\eeqa{\end{eqnarray}}
\def\LCDM{\ensuremath{\Lambda}CDM}

\def\head{ \vbox to 0pt{\vss \hbox to 0pt{\hskip 440pt\rm
      LA-UR-10-07069\hss} \vskip 25pt}}

\def \kms {\ifmmode  \,\rm km\,s^{-1} \else $\,\rm km\,s^{-1}  $ \fi }
\def \kpc {\ifmmode  {\,\rm kpc}  \else ${\rm  kpc}$ \fi  }  
\def \hkpc {\ifmmode  {h^{-1}\rm kpc}  \else ${h^{-1}\rm kpc}$ \fi  }  
\def \hMpc {\ifmmode  {h^{-1}\rm Mpc}  \else ${h^{-1}\rm Mpc}$ \fi  }  
\def \Mpch {\ifmmode  {h^{-1}\rm Mpc}  \else ${h^{-1}\rm Mpc}$ \fi  }  
\def \Msun {\ifmmode {\rm M}_{\astrosun} \else ${\rm M}_{\astrosun}$ \fi} 
\def \hMsun {\ifmmode h^{-1}\,\rm M_{\astrosun} \else $h^{-1}\,\rm M_{\astrosun}$ \fi}
\def \Gyr {\ifmmode\, \rm Gyr \else $\,$Gyr \fi}

\def \LCDM {\ifmmode \Lambda{\rm CDM} \else $\Lambda{\rm CDM}$ \fi}
\def \sig8 {\ifmmode \sigma_8 \else $\sigma_8$ \fi} 
\def \OmegaM {\ifmmode \Omega_{\rm m} \else $\Omega_{\rm m}$ \fi} 
\def \Omegab {\ifmmode \Omega_{\rm b} \else $\Omega_{\rm b}$ \fi} 
\def \OmegaL {\ifmmode \Omega_{\rm \Lambda} \else $\Omega_{\rm \Lambda}$\fi} 
\def \Deltavir {\ifmmode \Delta_{\rm vir} \else $\Delta_{\rm vir}$ \fi}
\def \rhocrit {\ifmmode \rho_{\rm crit} \else $\rho_{\rm crit}$ \fi}
\def \rhou {\ifmmode \rho_{\rm u} \else $\rho_{\rm u}$ \fi}
\def \zc {\ifmmode z_{\rm c} \else $z_{\rm c}$ \fi}

\title[The edge of galaxy formation II]{The edge of galaxy formation II:  evolution of Milky Way satellite analogues after infall}

\author[J. Frings et al.]{Jonas Frings$^{1,2}$\thanks{E-mail: frings@mpia.de},
Andrea Macci\`o$^{1,3}$\thanks{E-mail: maccio@nyu.edu},
Tobias Buck$^{1}$, Camilla Penzo$^{4,1}$, Aaron Dutton$^{3}$,
\newauthor{Marvin Blank$^{3,5}$, Aura Obreja$^{6,3}$}
\\
$^1$Max-Planck-Institut f\"ur Astronomie, K\"onigstuhl 17, 69117 Heidelberg, Germany\\
$^{2}$Astronomisches Recheninstitut, Zentrum f{\"u}r Astronomie der Universit{\"a}t Heidelberg, Philosophenweg 12, 69120 Heidelberg, Germany \\
$^{3}$New York University Abu Dhabi, PO Box 129188, Abu Dhabi, United Arab Emirates\\
$^{4}$Laboratoire\,Univers\,et\,Th\'eories,\,UMR\,8102\,CNRS,\,Observatoire\,de\,Paris,\,Universit\'e\,Paris Diderot,\,5 Place\,Jules\,Janssen,\,92190\,Meudon,\,France\\
$^{5}$Institut f\"{u}r Theoretische Physik und Astrophysik, Christian-Albrechts-Universit\"{a}t zu Kiel, Leibnizstr. 15, D-24118 Kiel, Germany\\
$^{6}$Universit\"ats-Sternwarte, Ludwig-Maximilians-Universit\"at M\"unchen, Scheinerstr. 1, D-81679 M\"unchen, Germany
}

\date{Accepted XXX. Received YYY; in original form ZZZ}

\pubyear{2017}

\begin{document}

\label{firstpage}
\pagerange{\pageref{firstpage}--\pageref{lastpage}}
\maketitle

\begin{abstract}
In the first paper we presented 27 hydrodynamical cosmological simulations
of galaxies with total masses between $5 \times 10^8$ and $10^{10}$ \Msun.
In this second paper we use  a subset of these cosmological simulations
as initial conditions (ICs) for  more than forty hydrodynamical simulations of satellite
and host galaxy interaction. Our cosmological ICs seem to suggest
that galaxies on these mass scales have very little rotational support
and are velocity dispersion ($\sigma$) dominated.
Accretion and environmental effects increase the scatter in the
galaxy scaling relations (e.g. size - velocity dispersion) in very
good agreement with observations. Star formation is substantially
quenched after accretion. Mass removal due to tidal forces 
has several effects: it creates a very flat stellar velocity dispersion profiles,
and it reduces the dark matter content at all scales (even in the centre), which
in turn lowers the stellar velocity on scales around 0.5 kpc even when the galaxy
does not lose stellar mass.
Satellites that start with a cored dark matter profile are more prone to either
be destroyed or to end up in a very dark matter poor galaxy. Finally, 
we found that tidal effects always increase the ``cuspyness'' of the dark matter profile,
even for haloes that infall with a core.
\end{abstract}

\begin{keywords}
cosmology: theory -- dark matter -- galaxies: formation -- galaxies: kinematics and dynamics -- methods: numerical
\end{keywords}

%%%%%%%%%%%%%%%%%%%%%%%%%%%%%%%%%%%%%%%%%%%%%%%%%%

%%%%%%%%%%%%%%%%% BODY OF PAPER %%%%%%%%%%%%%%%%%%
%%%%%%%%%%%%%%%%%%%%%%%%%%%%%%%%%%%%%%%%%%%%%%%%%%%
\section{Introduction}\label{sec:introduction}
%%%%%%%%%%%%%%%%%%%%%%%%%%%%%%%%%%%%%%%%%%%%%%%%%%%

The current model for the formation and evolution of the Universe predicts a hierarchical assembly of collapsed structures, with 
small, low mass  dark matter haloes forming first and then subsequently merging to form more massive structures \citep{White1978, Blumenthal1984}. 
In such a picture, the baryonic component (i.e. the gas) will initially fall into  the potential wells created by the dark matter haloes;
then it will begin to efficiently cool in the centre of the  overdensity, and eventually  lead to the formation of stars and galaxies. 

In the so-called standard model for cosmology, gravity is described by general relativity under the presence of a cosmological constant $\Lambda$ \citep{Riess1998, Perlmutter1999} 
and the matter component is dominated by Cold Dark Matter \citep{Peebles1984}. 
While on large scales ($>$Mpc) this model is very successful  in predicting the observed structure of the Universe, 
it has been often claimed to have some issues in reproducing observations concerning the low mass end of the galaxy population. 
Those challenges of the $\Lambda$CDM model on small scales became known as the missing satellites  problem \citep{Klypin1999, Moore1994}, 
the cusp-core tension \citep{Flores1994, Moore1994, Oh2015} and the too-big-to-fail problem \citep{Boylan-Kolchin2011}. 
Several of these issues arise in local galaxies and mostly in the satellites around the Milky Way, so investigating the low mass end (the edge) of galaxy formation,
i.e. galaxies with stellar masses below  $\approx 10^7\,\Msun$, can provide very important insights on the validity of our current cosmological model. 

However the physics involved in the process of structure formation becomes more and more complicated when going from large scales, that can be well described just by gravity, 
to small scales where baryonic effects like gas cooling, star formation and feedback play major roles. It has been pointed out that most if not all of the failures of $\Lambda$CDM can be alleviated 
when pure $N$-body simulations are replaced by more sophisticated hydrodynamical simulations which include all baryonic effects mentioned above \citep{Dutton2016}.

Hydrodynamical simulations usually divide into cosmological volume simulations \citep{Grand2017, Schaye2015, Vogelsberger2014, Sawala2016} and zoom-in simulations of single objects 
\citep{Maccio2012b, Stinson2013, Aumer2013, Hopkins2014, Marinacci2014, Wang2015, Dutton2016, Wetzel2016}. 
Nowadays it is possible to achieve resolutions of few million particles per object and  with a set of different zoom-in simulations the whole mass spectrum of galaxy formation can be covered 
\citep{Wang2015, Chan2015}. On the other hand interacting systems like the Milky Way and  its satellite galaxies, that differ by a factor $10^4$ in mass,
cannot be easily simulated; in fact to achieve a sufficient resolution in the satellite ($\approx 10^6$ particles) one would end up with $10^{10}$ particles in the Milky Way halo, 
which is far from manageable even for modern supercomputers. For comparison the simulation with the best mass resolution of a Milky Way system, the Latte project \citep{Wetzel2016}, has total of a few $10^7$ particles.

To overcome this issue different approaches have been suggested in the literature. The majority of studies have been made using modeled (pre-cooked) galaxies and then studying their evolution 
on (several) orbits around their host in isolated simulations \citep{Kazantzidis2004, Mayer2006, Kang2008, Donghia2009, Chang2013, Kazantzidis2017}. 
While this may be a good approach to investigate the second part of the evolution of a satellite galaxy, the interaction of the satellite with its host, 
the use of these modeled galaxies neglects the first part of the life of satellite: its formation and evolution before the accretion onto the host. 

In this series of papers we have decided to use a new approach, which aims to combine the insights from full cosmological hydrodynamical simulations with the very high resolution
attainable in simulations of binary galaxy interactions.
Namely we use cosmological hydrodynamic simulations to produce  realistic initial conditions for the isolated simulation of satellite-host interaction. 
In the first paper  \citep{maccio2017} (from now on referred to as PaperI) we have introduced our cosmological simulations and presented a detailed analysis of the properties
of our simulated galaxies {\it before} the accretion. In this second paper (PaperII) we study the evolution of these galaxies after they have been exposed to the environmental 
effects of their central object. Our final goal is to study the effects of accretion and environment on realistic satellite galaxies.

In this paper we start in section 2 with a short description of the simulation code, how the isolated accretion simulations are set up and how we model different physical effects. 
In section 3 we present the results of our simulations, focussing on mass losses, scaling relations, stellar kinematics and dark matter structure. 
Finally, in section 4 we present our discussion and conclusion on the environmental effects on realistic satellite galaxies in a Milky Way mass halo.

%%%%%%%%%%%%%%%%%%%%%%%%%%%%%%%%%%%%%%%%%%%%%%%%%%%
\section{Simulations} \label{sec:simulations}
%%%%%%%%%%%%%%%%%%%%%%%%%%%%%%%%%%%%%%%%%%%%%%%%%%%
\subsection{Cosmological simulations} \label{sec:cosmosims}
We use a subsample of seven simulations of the dwarf galaxy sample introduced in PaperI.
The original sample contains 27 cosmological zoom-in simulations of central haloes in the mass range of $5\times10^8 < M_{\mathrm{dark}} / \Msun < 2\times10^{10}$ of which 19 
form a galaxy in their centre.  
The cosmological simulations were run using the smoothed particle hydrodynamics code {\sc gasoline}  \citep{wadsley2004} until redshift $z=1$ in a $\Lambda$CDM cosmology using the WMAP 7 set of cosmological parameters \citep{komatsu2011}: Hubble parameter $H_0$= 70.2 \kms Mpc$^{-1}$, matter density $\Omega_\mathrm{m}=0.2748$, dark energy density
$\Omega_{\Lambda}=1-\Omega_\mathrm{m} -\Omega_\mathrm{r}=0.7252$, baryon density
$\Omega_\mathrm{b}=0.04572$, normalization of the power spectrum $\sigma_8 = 0.816$, slope of the inital power spectrum $n=0.968$.

The code set up was the same as for the MaGICC project \citep{Stinson2013, Kannan2014, Penzo2014} and included metal cooling, chemical enrichment, star formation and feedback from supernovae (SN) and massive stars (the so-called Early Stellar Feedback). The density threshold for star formation is set to $60\,\mathrm{cm}^{-3}$ which represents the mass of a smoothing kernel (32 particles) in a sphere of radius of the softening \citep[see][for more details]{Wang2015}, while the star formation efficiency is set to $c_\star=0.1$. The cooling function includes the contribution of metals as described 
in \citet{Shen2010} and we also include photoionisation and heating from the ultraviolet background following \citet{Haardt2012} and  Compton cooling. 
The SN feedback relies on the blast-wave recipe described in \citet{Stinson2006}. Finally we identified the haloes using the amiga halo finder  ({\sc ahf}\footnote{http://popia.ft.uam.es/AMIGA}) \citep{ahf}.

The mass resolution of the zoom-in region of the simulations is shown in Table \ref{tab:sample}. Gas particles have an inital mass of $m_\mathrm{gas,init}=f_\mathrm{bar}\cdot m_\mathrm{DM}$ while stellar particle start with inital masses $m_\mathrm{star, init}=\frac{1}{3}m_\mathrm{gas,init}$, where $f_\mathrm{bar}=\frac{\Omega_\mathrm{b}}{\Omega_\mathrm{dark}}$ is the cosmic baryon fraction.

%%%%%%%%%%%%%%%%%%%%%%%%%%%%%%%%%%%%%%%%%%%%%%%%%%%
\begin{table}
\centering
\caption{Virial dark matter mass, stellar mass, virial radius (defined by an overdensity of $200\,\rho_\mathrm{crit}$) and dark matter particle mass  of the selected subsample of simulations. 
For all simulations the gravitational softening for the dark matter, stellar and gas particles is $\varepsilon_\mathrm{DM}=31\,\mathrm{pc}$ and $\varepsilon_\mathrm{gas}=\varepsilon_\mathrm{star}=14\,\mathrm{pc}$, respectively.}
\label{tab:sample}
\begin{tabular}{lccccr}

Name& $M_\mathrm{DM}[\Msun]$ & $M_\mathrm{star}[\Msun]$  & $r_{200}[\kpc]$ &$N_\mathrm{star}$& $m_\mathrm{DM}[\Msun]$\\ \hline\hline
satI&	1.02e+10 &	8.97e+06&	31.34 & 81629& 1.36e+03 \\ \hline	
satII&	5.52e+09&	1.81e+06&	25.65&16822&1.36e+03\\ \hline	
satIII&		5.61e+09&	1.20e+06&	25.60&6902&2.02e+03\\ \hline	
satIV&	2.92e+09&	5.46e+05&	20.65&5202&1.36e+03 \\ \hline	
satV&	4.49e+08&	4.25e+04&	10.97&408&1.36e+03 \\ \hline	
darkI&	3.04e+09&	0&			20.93&0&2.02e+03\\ \hline	
darkII&	2.81e+09&	0&			20.32&0&1.36e+03\\
\end{tabular}
\end{table}
%%%%%%%%%%%%%%%%%%%%%%%%%%%%%%%%%%%%%%%%%%%%%%%%%%%

\subsection{Satellite initial conditions}
\label{sec:ics}
Starting from the redshift $z=1$ snapshots of the cosmological simulations described in section \ref{sec:cosmosims} we cut out seven halos  and their surrounding structures up to a distance of four virial radii from their centre (for the virial radius we used the region  enclosing a density equal to 200 times $\rho_\mathrm{crit}$, where $\rho_\mathrm{crit}$ is the cosmic critical matter density).
These cut out regions were then transformed from  cosmological (i.e. expanding) coordinates to physical ones and used as initial conditions for our subsequent accretion simulations.
Table \ref{tab:sample} contains the main parameters of our selected haloes, the name {\it sat} is used for halos that contained stars at the starting redshift ($z=1$) while we reserve the name {\it dark}
for haloes without stars.
As a first step we run all the haloes (dark and luminous) ``in isolation'' from $z=1$ to $z=0$, meaning we evolved them without the presence of the central halo, in order to have a base line for the evolution of our galaxies; we will refer to this set of simulations as the  \textit{isolated} runs.

\subsection{Central object parametrization}
\label{sec:central}

We used two different models for the parametrization of the central object. 
At first we described it as an analytical potential  consisting of the superposition of two distinct potentials for the dark matter and the stellar disc.
For the dark matter halo we used a Navarro-Frenk-White (NFW) potential \citep{nfw1996} with a mass $M_{200}=1\times10^{12}\,\Msun$, a concentration parameter $c=10$
\citep{Dutton2014}, and a virial radius $r_{200}=210\kpc$. For the stellar body we used  a Miyamoto \& Nagai potential \citep{miyamoto1975} with a disc mass $M_\mathrm{disc}=5\times10^{10}\,\Msun$, disc scale length $R_\mathrm{disc}=3.0\kpc$ and height $h_\mathrm{disc}=0.3\kpc$. The disc is aligned with the $x-y$ plane of the simulation. 
 
The use of an analytic potential makes the simulation faster but misses one possible important ingredient: dynamical friction. 
In order to estimate its effect we also used a live halo (i.e. made with particles) without a disc component. 

To construct the (central) galaxy model we apply
the method described in \cite{Springel2005b} and in \cite{Moster2014}. The halo is described by a NFW density profile with the same parameters used for
our analytic potential (a virial radius of $r_{200}=210\kpc$, a virial mass of $M_{200}\approx 1\times10^{12} \Msun$ and a concentration parameter of $c=10$).
We used two resolution levels for this live halo with $10^5$ and $10^7$ particles, respectively. 
Runs performed with the live halo are discussed in the next section.

\subsection{Orbits}
\label{sec:orbits}

We select six different orbits and run the simulations until redshift $z=0$. 

All orbits start at the virial radius at the coordinates (x, y, z)$=(210.0,\,0,\,0)\kpc$ but they differ in the initial velocity of the satellite and in the angle between the plane of the orbit and the stellar disc
of the host halo. The parameters of all orbits are summarized in table \ref{tab:orbits}, where we have ordered the six orbits by their  ``disruptiveness'', i.e. \textit{orbitI} is the most gentle orbit, causing the least deviations from the isolated run (for example in mass loss) while  \textit{orbitV} provides the most violent interaction between the satellite and the central object, with the exception of the complete radial infall.
The pericentre  distance is shown only for \textit{satV}. We want to point out that compared to the orbits of surviving Milky Way satellites even \textit{orbitI} with a pericentre of $25\kpc$ is quite extreme. \citet{Garrison2017} show that in their simulations only $5\,\%$ of the surviving satellites around Milky Way mass galaxies have orbital pericentres below $20\kpc$. The choice for such strong orbits has been dictated by our aim to braket the possible scenarions between unperturbed evolution (the isolation case) and strong interactions. 

\begin{figure}
\includegraphics[width=0.47\textwidth]{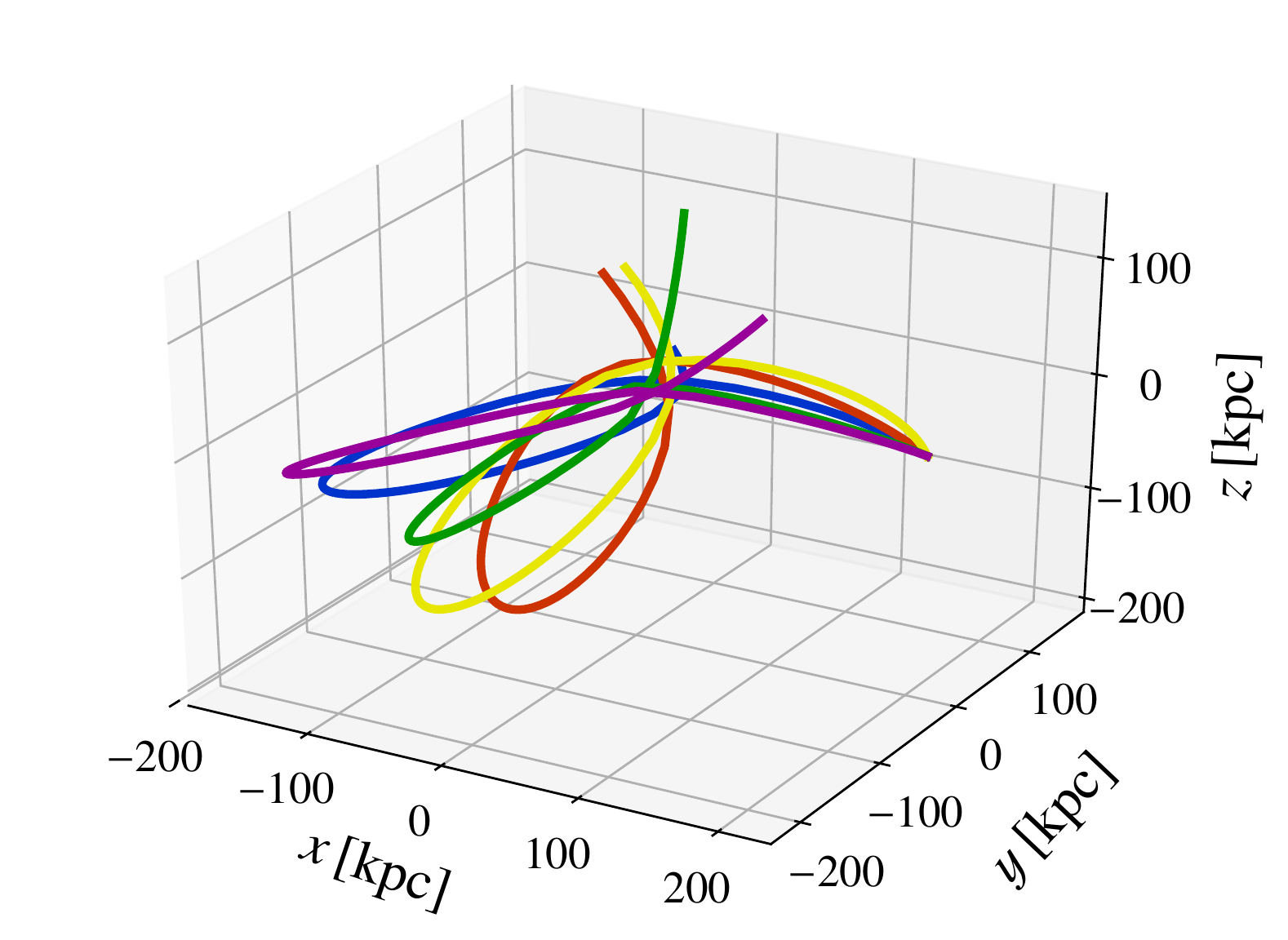}
\vspace{-.35cm}
\caption{Visualization of the orbits (\textit{orbitI} to \textit{orbitV}) presented in Table \ref{tab:orbits} from infall to $5.7\Gyr$ after infall. The colour coding is the same as in Table \ref{tab:orbits}.}
\label{fig:orb3D}
\end{figure}

In Fig. \ref{fig:orb3D} we show the trajectory of \textit{satV} on  the orbits from \textit{I} to \textit{V} from $z=1$ to $z=0$ (5.7 Gyr). The colours correspond to individual orbits as introduced in Table \ref{tab:orbits}.
The centre of each satellite is defined as the position of the maximum of the stellar (dark matter)  density distribution for the luminous (dark) satellites. The position of this maximum is derived via a shrinking spheres algorithm according to \citet{power2003}.

\begin{figure}
\includegraphics[width=0.47\textwidth]{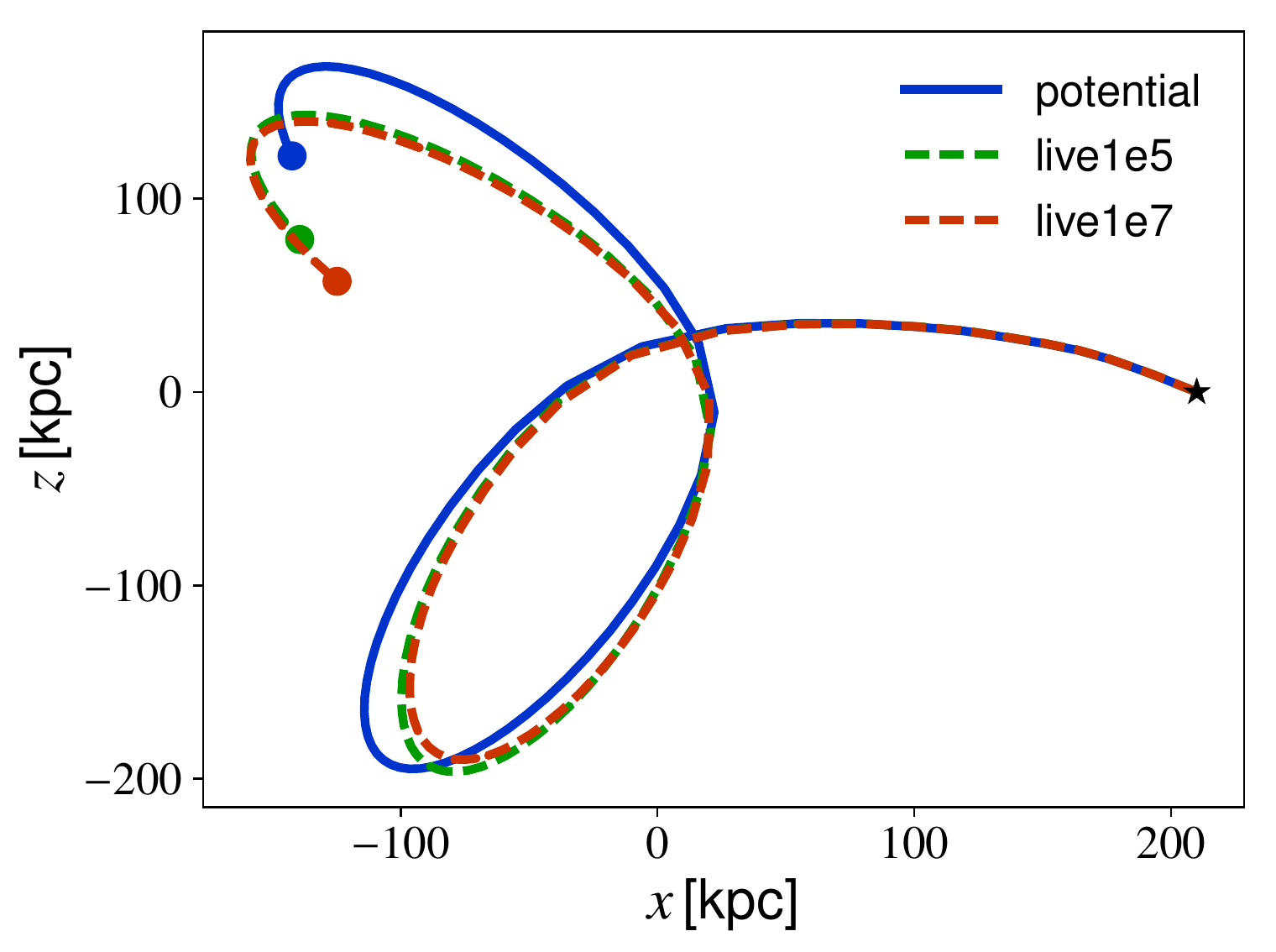}
\vspace{-.35cm}
\caption{Projection of \textit{orbitII} onto the $x-z$ plane comparing the evolution in an analytic potential (blue), in a live halo with $10^5$ (green) and $10^7$ (red) dark matter particles. A star marks the start of the orbit while the dot marks the end.}
\label{fig:orb2D}
\end{figure}

In Fig. \ref{fig:orb2D} we compare the orbit evolution of \textit{orbitII} in the analytic potential
in the live halo at two different resolutions levels:  $10^5$ and $10^7$ particles. Dynamical friction does slightly modify the orbit, but the effect is quite small
and independent of the resolution of the live halo, the same result holds also for the other orbits.
Since our choice of orbits has been practically random it is fare to say that the small effect of dynamical friction is very similar to a slightly different choice of orbits, and can 
be then neglected in our study.

\subsection{Ram pressure}
\label{sec:ram}

Even with a live halo our set up is not able to take into account the effect of ram pressure between the (hot) gas in the host halo and the gas in the satellite.
We therefore add an analytic recipe for ram pressure to our simulations. 
At every basic time-step (1$\,\mathrm{Myr}$), we remove all gas particles within the satellite that are below a certain threshold density $\rho_\mathrm{th}$. \\
This density threshold evolves with time according to the following expression:
\begin{equation}
\label{eq_ram}
\rho_\mathrm{th}(t)=\begin{cases}
					\rho_\mathrm{max}\,\left[\frac{\rho_\mathrm{min}}{\rho_\mathrm{max}} \right]^{\left(1-\frac{t}{\tau_\mathrm{ram}}\right)^a}, &\mbox{for}\:t<\tau \\
			\rho_\mathrm{max}, &\mbox{for}\:t>\tau
					\end{cases}
\end{equation}

where $\rho_\mathrm{min}$ is the gas density at four virial radii; $\rho_\mathrm{max}$ is the gas density in the centre of the halo; $\tau_\mathrm{ram}$ is the gas removal time scale (see later for more details);  $a$ is a free paramter and $t$ denotes the runtime of the simulation. 

If in equation \eqref{eq_ram} we set $\rho_\mathrm{th}(t=0)=\rho_\mathrm{min}$ and $\rho_\mathrm{th}(t=\tau_\mathrm{ram})=\rho_\mathrm{max}$, this implies that all gas will be removed
(outside in) at time  $t=\tau_\mathrm{ram}$. We fixed the value of the parameter $a$ at 0.2, since this ensures that $\rho_\mathrm{th}$ follows the radial profile of the gas density.
Finally we set the gas removal time scale $\tau_\mathrm{ram}$  approximately equal to a half the dynamical time scale of the orbit,  so that all gas is removed after one 
pericentre passage (see Table \ref{tab:orbits}).

%The evolution of $\rho_\mathrm{th}$ as described in \eqref{eq_ram} with $\rho_\mathrm{th}(t=0)=\rho_\mathrm{min}$ and $\rho_\mathrm{th}(t=\tau_\mathrm{ram})=\rho_\mathrm{max}$ implies a removal of gas starting at the outskirts and moving radially inwards until at $t=\tau_\mathrm{ram}$ all gas is removed. We chose $a=0.2$ to make sure that $\rho_\mathrm{th}$ follows the approximate radial gas density profile inwards and linearly in time. 
%The gas removal time scale $\tau_\mathrm{ram}$ we set for each orbit approximately to a half the dynamical time scale of the orbit so that all gas is removed after one pericentre passage (see Table \ref{tab:orbits}). 

We do not expect ram pressure to be important from a dynamical point of view, since our galaxies are very strongly dark matter dominated (see PaperI), but it might be important to
accelerate the galaxy quenching. On the other hand by comparing \textit{satIV} on \textit{orbitII} with and without ram pressure we found no strong changes
in its star formation rate, with both set ups showing a very similar quenching behaviour compared to the isolated run. {}

Even if we do not see a large difference in the outcomes of the simulations with and without the ram pressure model, we still apply the ram pressure model to all simulations
since it decreases the computational cost (by reducing the number of gas particles within the galaxy).

%Since the galaxies are mainly dominated by dark matter the removal of the gas has not a very large effect on the galaxies except for the quenching of star formation. However, a run of \textit{satIV} on \textit{orbitII} without the ram pressure model has shown the same quenching behaviour compared to the isolated run.

%%%%%%%%%%%%%%%%%%%%%%%%%%%%%%%%%%%%%%%%%%%%%%%%%%%
\begin{table}
\centering
\caption{Compilation of the different orbit scenarios with their inital velocity, pericentre distance and orbit inclination with respect to the host galaxy disc. All orbit initiate at the coordinates $x=210\kpc$, $y=z=0$. The colour indicates the colour coding in Fig. \ref{fig:orb3D}.}
\label{tab:orbits}
\begin{tabular}{lcccr}

Name& $(v_x,\,v_y,\,v_z) \,[v_{200}]$  &$\tau_\mathrm{ram}\,[\Gyr]$& $r_\mathrm{min}\,[\kpc]$ &$\vartheta \,[\deg]$\\ \hline\hline
{\color{color1}\textit{orbitI}}&	$(-0.45,\,0.3,\,0)$&	1.5&25.52&0\\ \hline	
{\color{color2}\textit{orbitII}}&$(-0.45,\,0,\,0.3)$&	1.5&25.46&90\\ \hline	
{\color{color3}\textit{orbitIII}}&$(-0.2,\,0.2,\,0.2)$&1.5&25.26&45\\ \hline	
{\color{color4}\textit{orbitIV}}&$(-0.5,\,0,\,0.1)$&	1.4&7.94&90\\ \hline	
{\color{color5}\textit{orbitV}}&	$(-0.5,\,0.1,\,0)$&	1.4&7.2&0\\ \hline	
\textit{radial}&	$(-0.5,\,0,\,0)$&1.3&-&0\\
\end{tabular}
\end{table}
%%%%%%%%%%%%%%%%%%%%%%%%%%%%%%%%%%%%%%%%%%%%%%%%%%%


\section{Results}
\label{sec:results}

\subsection{Rotational support}
\label{sec:rotsup}

%%%%%%%%%%%%%%%%%%%%%%%%%%%%%%%%%%%%%%%%%%%%%%%
\begin{figure}
\includegraphics[width=0.47\textwidth]{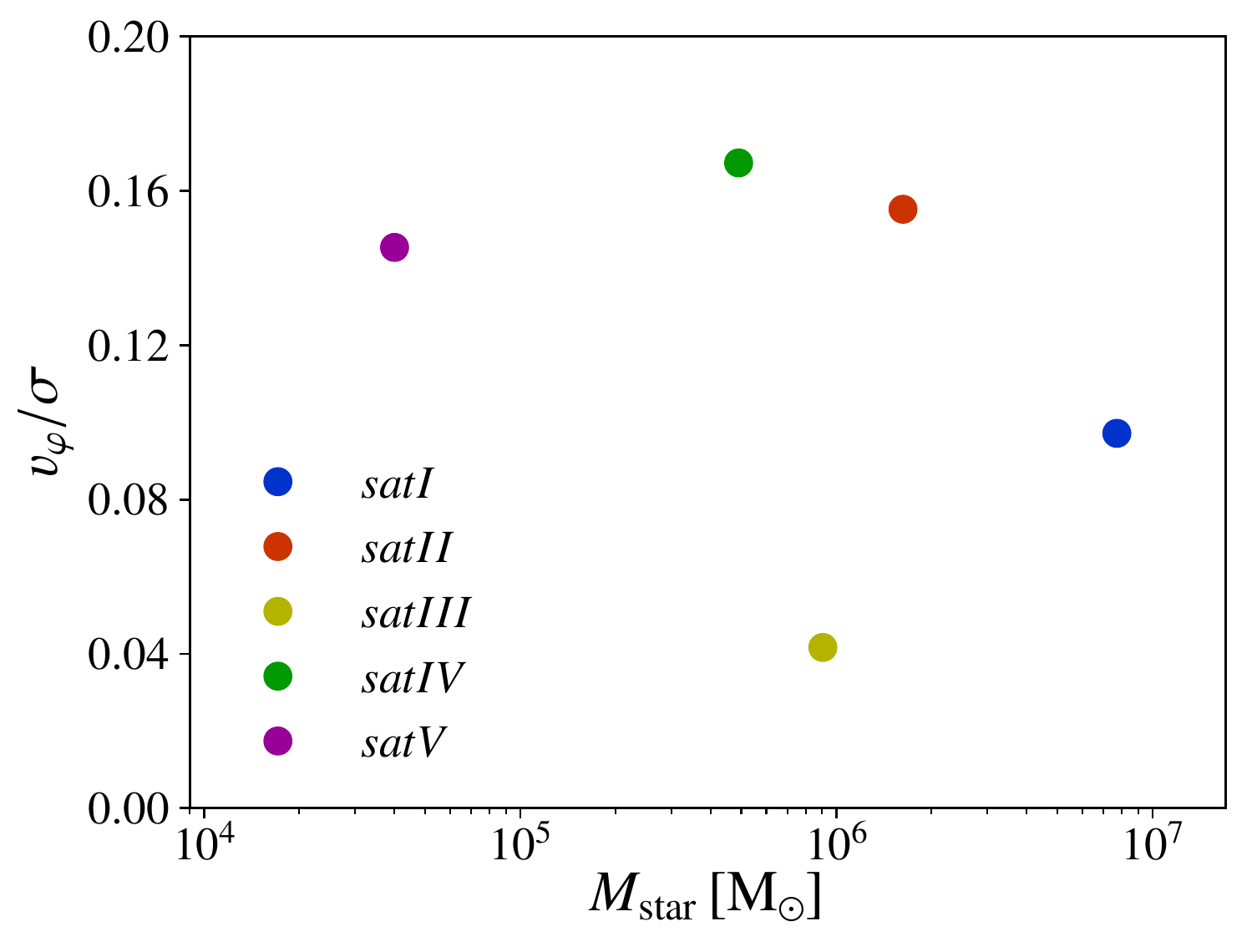}
\vspace{-.35cm}
\caption{Amount of rotation along the axis of total stellar angular momentum of the stars compared to stellar velocity dispersion as a function of stellar mass.}
\label{fig:vrot}
\end{figure}
%%%%%%%%%%%%%%%%%%%%%%%%%%%%%%%%%%%%%%%%%%%%%%%

One main difference between our approach and previous studies in the literature is the use of cosmological simulations as initial conditions, 
it is then interesting to check the dynamical state of our galaxies before the infall.
In Fig. \ref{fig:vrot} we show the amount of rotational support of the stellar component in the dwarf galaxies at $z=1$.  
We obtain the rotational velocity $v_\varphi$ by averaging the individual velocities of the stellar particles  in $\vec{e}_\varphi$ direction. 
The unit vector $\vec{e}_\varphi$ is set as the circumferential direction with respect to the axis defined by the  total stellar angular momentum of stars within the half mass radius. 
The velocity dispersion  $\sigma$ is simply given  by $\sigma=\sigma_{3D}/\sqrt{3}$, where $\sigma_{3D}$ is the three dimensional velocity dispersion of the stellar particles in the half mass radius.
As shown in Fig. \ref{fig:vrot} our galaxies show that there is not much rotational support and their structure can be described by a single isotropic component with practically no signs of a stellar disc
even before infall. This is in agreement with previous studies which showed that in cosmological simulations isolated dwarf galaxies as well as satellite galaxies at the low mass end seem to be dispersion supported systems \citep{Wheeler2017}.

However this is quite different from several previous works studying satellite-host interaction, which usually adopted values of $v_\mathrm{rot}/\sigma \approx 2$ \citep{Kazantzidis2017} with a well defined disc component, and then 
witness a ``morphological transformation'' within the host halo \citep{Lokas2010}. In our case no morphological transformation is needed since
cosmological simulations seem to indicate that galaxies, on our mass scales, are already quite ``messy'' and do not show the presence of a stellar disc.

\subsection{Environmental effects on galaxy properties}
\label{sec:gal_properties}

All the satellites survive till redshift $z=0$ on orbits from \textit{orbitI} to \textit{orbitV} with the exception of \textit{satI} on \textit{orbitV}, since in this case our centering algorithm is
not able to find a well defined stellar (or dark matter) centre for the satellite, as it is also confirmed by  a visual inspection which shows the satellite being completely 
destroyed. The same happens for the radial orbit scenario, in which all satellites are destroyed with no exceptions, we plan to analyze these tidal debris in a forthcoming paper.
Since in this paper we are interested in the properties of ``alive'' satellites at $z=0$, 
we will therefore focus on  \textit{orbitI} to \textit{orbitV}  for all our satellites but excluding \textit{orbitV} for  \textit{satI}.

\subsubsection*{Mass loss and abundance matching}
\label{sec:mass_loss}
%%%%%%%%%%%%%%%%%%%%%%%%%%%%%%%%%%%%%%%%%%%%%%%
\begin{figure}
\includegraphics[width=0.47\textwidth]{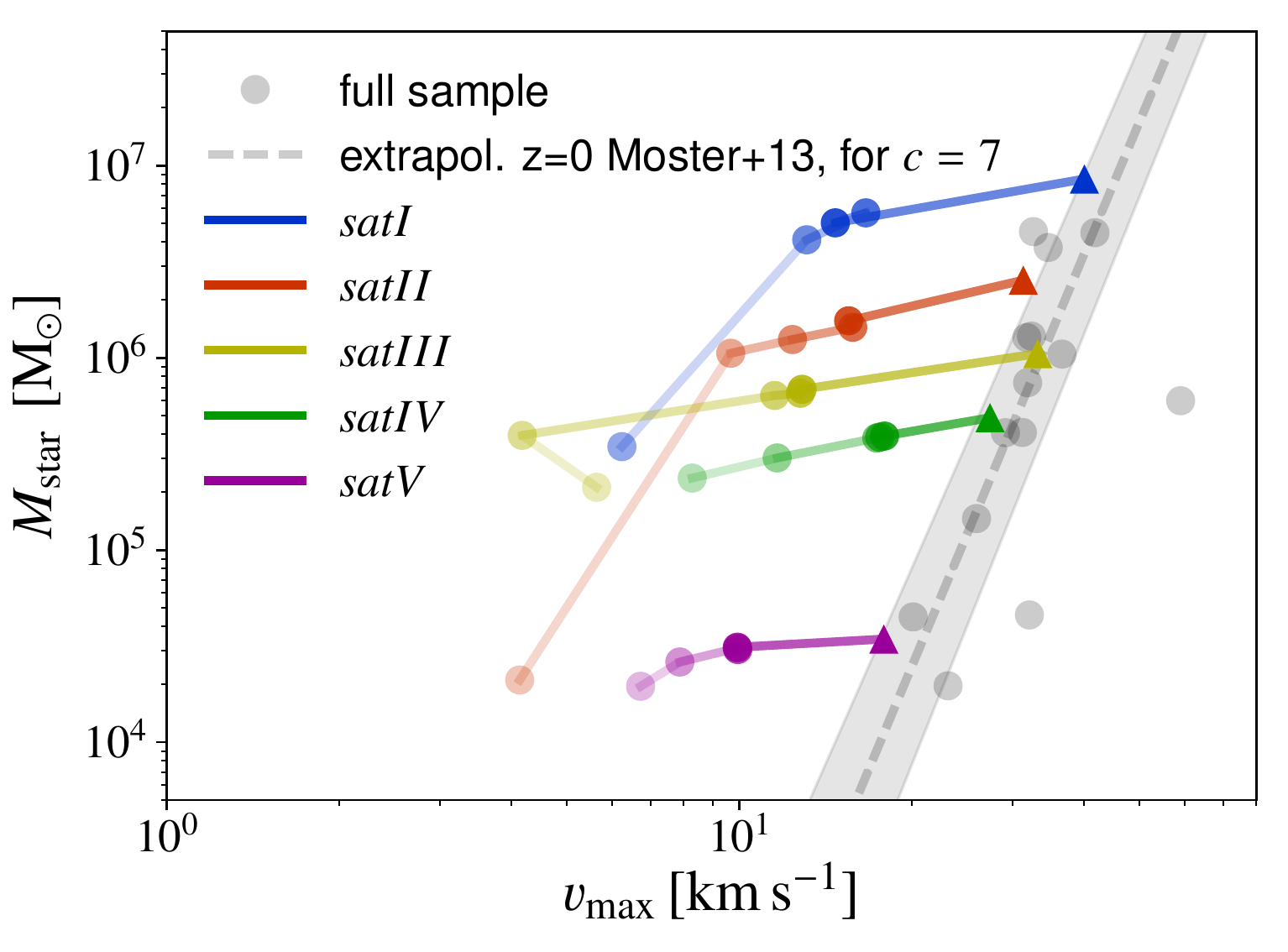}
\vspace{-.35cm}
\caption{The stellar mass within three stellar 3D half mass radii (at infall) as a function of the maxium circular velocity. The gray band shows the Moster relation and its errors translated to a function of $v_\mathrm{max}$ for a concentration of $c=7$. Triangles denote the isolated simulations while the filled circles denote the different orbits. The more violent the orbit, the fainter is the colour of the dots.}
\label{fig:moster}
\end{figure}
%%%%%%%%%%%%%%%%%%%%%%%%%%%%%%%%%%%%%%%%%%%%%%%

In Fig. \ref{fig:moster} we show the stellar mass within three half mass radii at infall as a function of the maximum of the total circular velocity profile $v_\mathrm{max}$. 
The half light radius is determined by the radius of a sphere around the centre of the satellite containing half of its  stellar mass, we will refer to this measure as the
$3D$ half mass radius.
The (coloured) triangles mark the results for the isolated runs at redshift zero. The (coloured) filled circles represent the runs in the disc+halo potential, same colours refer to the same
satellite (they are also connected by a line to facilitate the comparison), while the strength of the colour goes from dark to faint as the orbit becomes more destructive, i.e. from orbits (\textit{orbitI} to \textit{orbitV}).
We will use the same colour scheme in the rest of the paper. Finally the grey circles represent the full sample of haloes presented in PaperI at $z=1$ and are only added for comparison.

The dashed grey line shows the extrapolation to low mass haloes of the abundance matching relation from \citep{moster2013} and its error band. 
Since for satellites it is hard to define the total halo mass we have translated this last quantity into a maximum circular velocity. This has been assuming 
a NFW potential for the total matter distribution  with concentration $c=7$ (which is the average concentration of our simulated haloes) 
and also assuming that the maximum circular velocity occurs at the radius $r=2.16\,r_s$ \citep{Bullock2001}
where $r_s$ is the NFW scale radius.  

Our galaxies started on the abundance matching relation (grey circles, see also Paper I), and they remain there when run in isolation (coloured triangles). 
Then depending on the orbit, they leave the relation as a consequence of tidal stripping. For quiet orbits they move almost parallel to the x-axis (i.e. with constant stellar mass), meaning
that the central region of the satellite is fairly unaltered, then for more disruptive orbits also the stellar component is affected and the stellar mass can shrink to up to $1\%$ of its initial value.
Overall the satellites seem to perform a characteristic curve in the stellar mass vs. maximum circular velocity plane due to tidal stripping.

\begin{figure}
\includegraphics[width=0.45\textwidth]{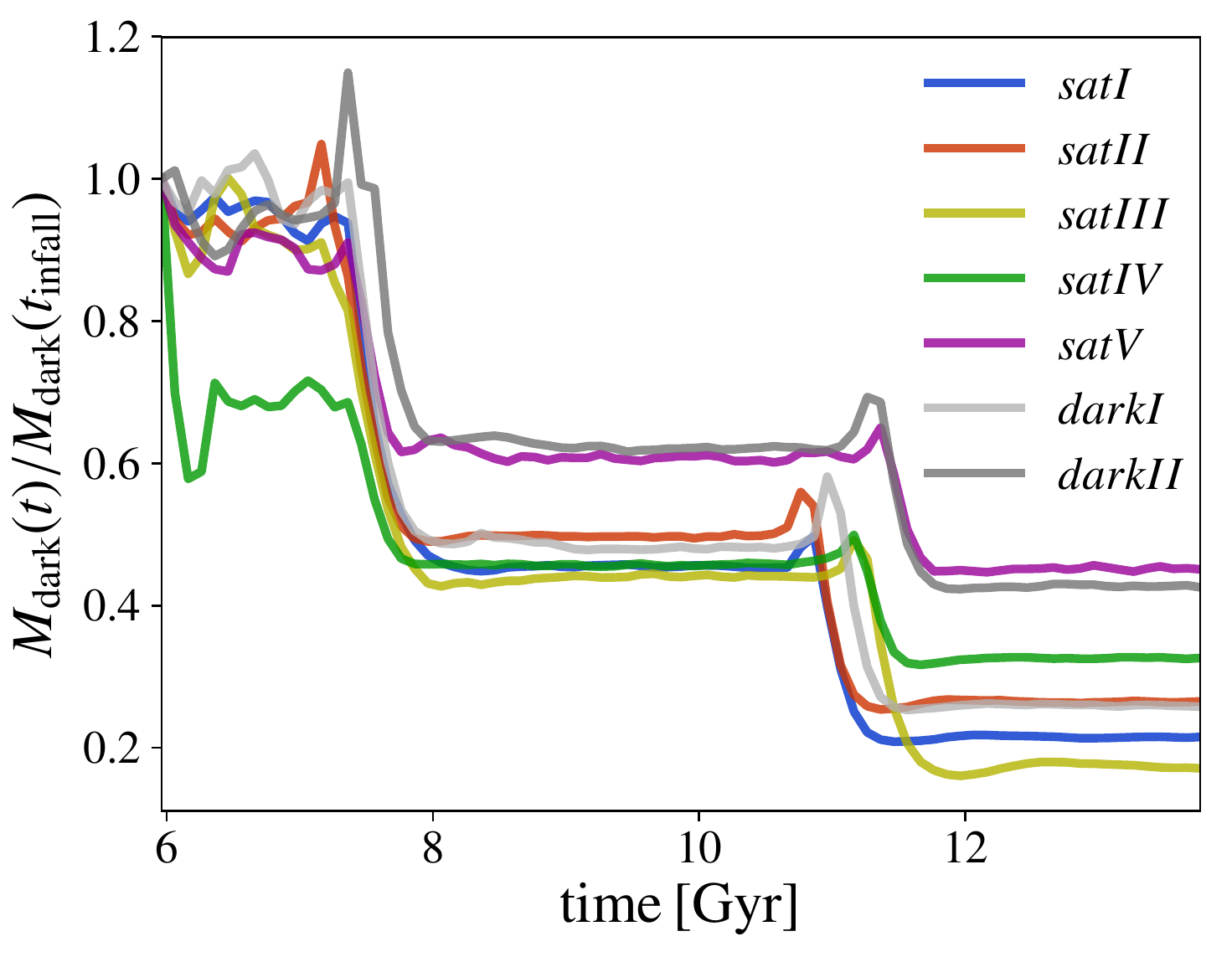}
\vspace{-.35cm}
\caption{Evolution of the dark matter mass in terms of the dark matter mass at infall of the seven satellites on \textit{orbitII}. Only the mass in $10\%$ of the virial radius at infall is considered.}
\label{fig:dm_mass}
\end{figure}
%%%%%%%%%%%%%%%%%%%%%%%%%%%%%%%%%%%%%%%%%%%%%%%

%%%%%%%%%%%%%%%%%%%%%%%%%%%%%%%%%%%%%%%%%%%%%%
\begin{figure}
\includegraphics[width=0.45\textwidth]{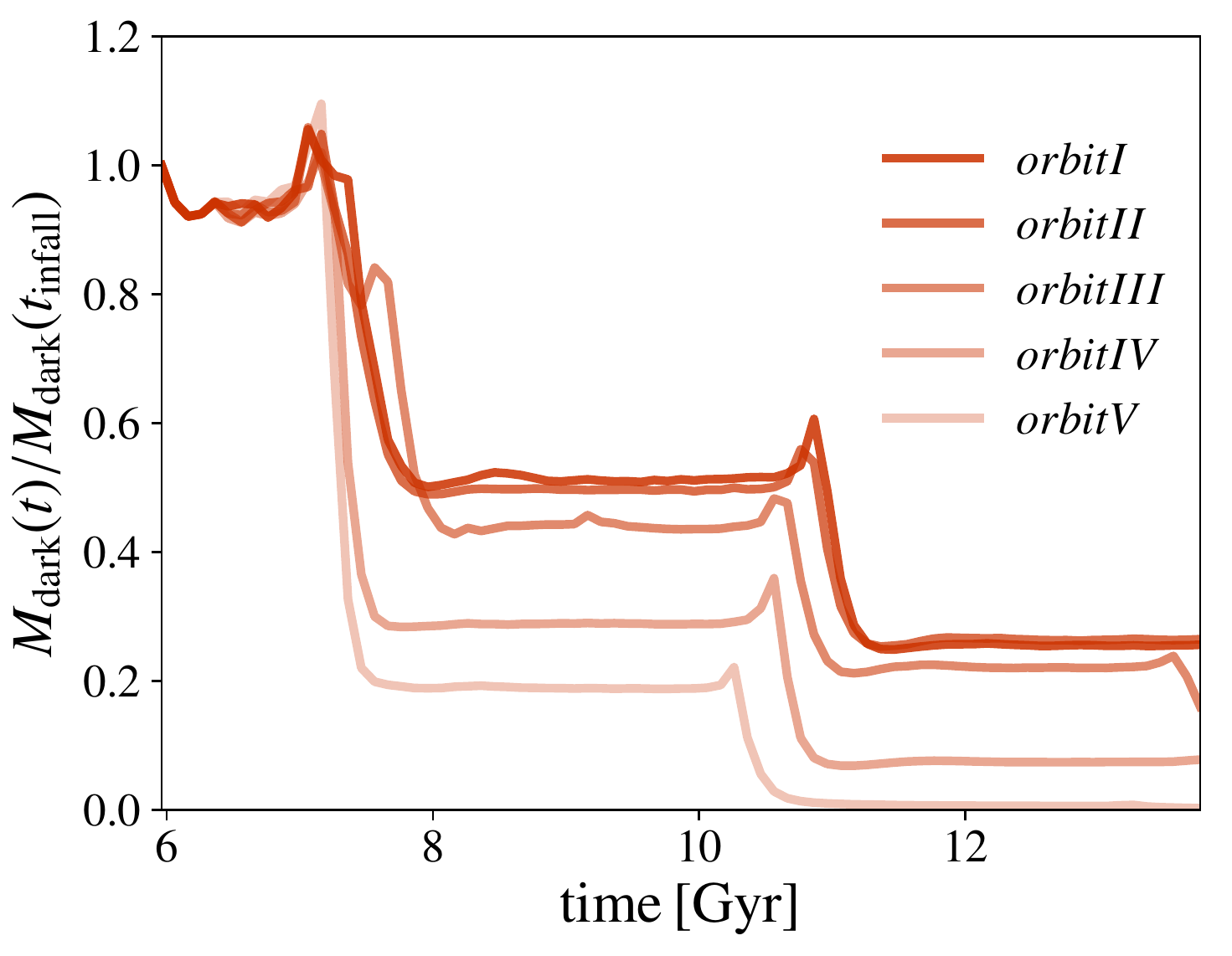}
\vspace{-.35cm}
\caption{Evolution of the dark matter mass in terms of the dark matter mass at infall for \textit{satII} on the five orbits. Only the mass in $10\%$ of the virial radius at infall is considered.}
\label{fig:dm_mass_orb}
\end{figure}
%%%%%%%%%%%%%%%%%%%%%%%%%%%%%%%%%%%%%%%%%%%%%%%

It is interesting also to look at the time evolution of the dark matter mass near the (luminous) centre of our haloes.
In Fig. \ref{fig:dm_mass} we show the evolution of the dark matter mass enclosed in $10\%$ of the initial (at infall) virial radius for all our 
satellites on the same orbit, namely \textit{orbitII}. 

Until the first pericentre passage at $t \approx 8\Gyr$ the dark matter mass is more or less stable with the exception of \textit{satIV} in which a (sub)substructure that passed nearby the centre led to an overestimation of the initial enclosed mass. 
During the first pericentre passage the satellites lose about half their inner dark matter mass. Then we find again a quite stable period untill the 
 second pericentre passage at $t\approx 11\Gyr$, when the satellites are stripped again and left with $15-50\%$ of their initial dark matter mass at redshift $z=0$.

%{\bf Andrea: this part is not clear to me, what did you want to say?} Even without the effect of dynamic friction one can see a slight trend of decay of the orbit with higher satellite mass. This has to be an effect of energy and orbital angular momentum redistribution between remnant and stripped material during the violent pericentre passage. In the end, also the total mass loss seems to correlate with intital mass, since the more massive satellites are stripped to $15\%$ while the smaller satellites end up with still close to $50\%$ of thier inital mass.

The second stripping event, corresponding to the second pericentre passage, occurs in the time range 11 to 12 \Gyr for all the satellites. 
The more massive satellites however seem to experience this event earlier (shortly after 11 \Gyr) than the less massive ones. 
This implies a sort of "decay" of the orbital trajectory for more massive satellites even in the absence of dynamical friction. This can be ascribed to the 
different redistribution of energy and orbital angular momentum between stripped material and the satellite remnant. 
%If this is the case, the stripped material would end up on orbits with higher energies and angular momenta than the satellite remnants orbit.
Finally in Fig. \ref{fig:dm_mass_orb} we show the evolution of the dark matter mass this time for a single satellite (\textit{satII}) on all five orbits. The fraction of dark matter remainig in the halo at redshift $z=0$ varies from $30 \,\%$ on \textit{orbitI} to just a few percent on \textit{orbitV}.

\subsubsection*{Stellar mass assembly}
\label{sec:stellar_hist}

%%%%%%%%%%%%%%%%%%%%%%%%%%%%%%%%%%%%%%%%%%%%%%%
\begin{figure}
\includegraphics[width=0.47\textwidth]{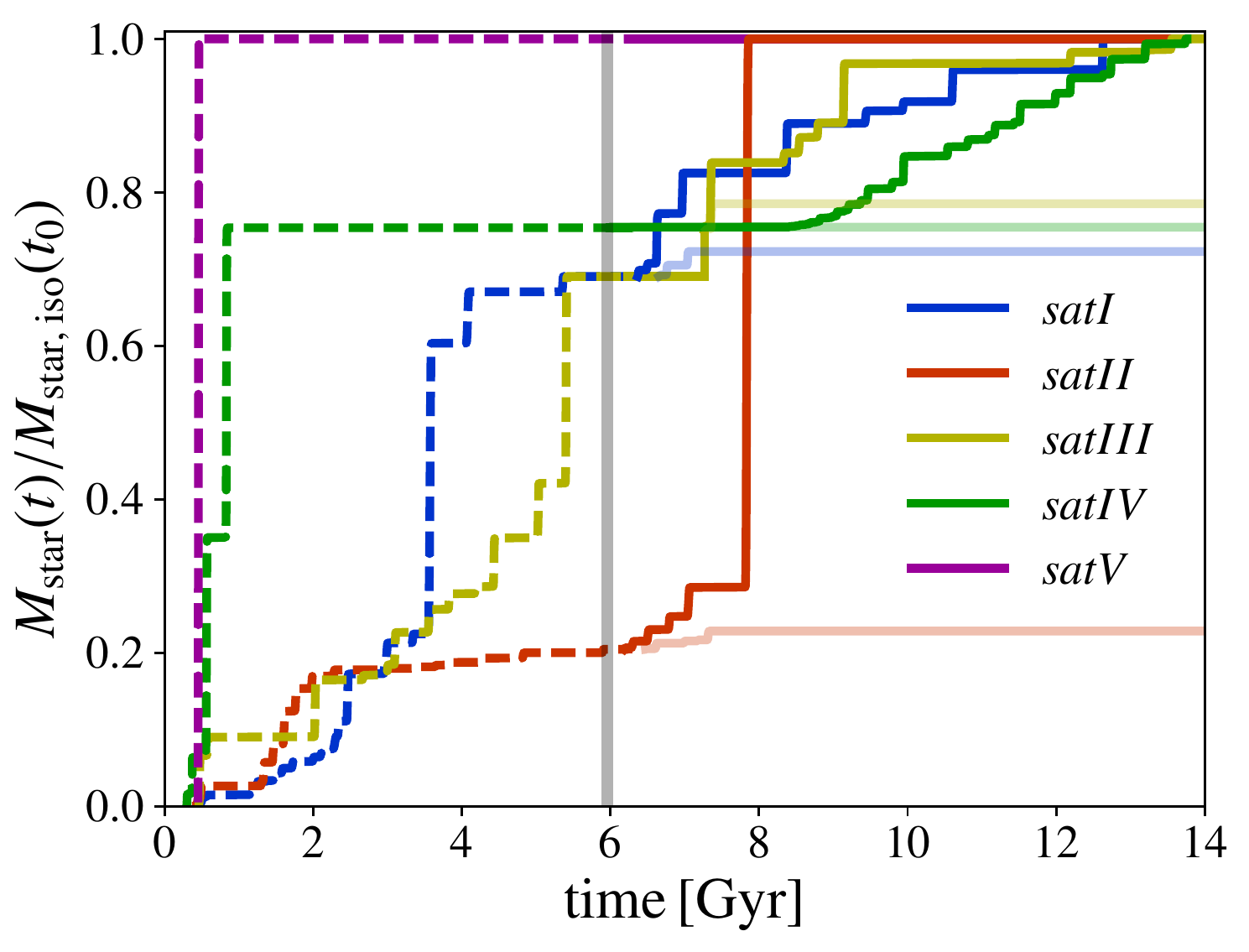}
\vspace{-.35cm}
\caption{Evolution of the stellar mass of the 5 luminous satellites before (dashed) and after (solid) the infall time ($z=1$). 
Isolated runs are shown with darker colours while orbits (specifically \textit{orbitII}) are shown with fainter ones. 
The stellar mass is shown in terms of the stellar mass at time $z=0$.}
\label{fig:stellar_mass}
\end{figure}
%%%%%%%%%%%%%%%%%%%%%%%%%%%%%%%%%%%%%%%%%%%%%%%

We now turn our attention to the luminous part of the satellites.
The evolution of the stellar mass in the cosmolocigal simulations until redshift $z=1$ (dashed) and physical simulations (solid) is shown in Fig. \ref{fig:stellar_mass} for the isolated run and \textit{orbitII} (faint lines). 

The stellar mass as a function of time is reconstructed from the formation times of the individual stellar particles. All stellar particles that remain within $10\%$ of the virial radius $r_\mathrm{200,infall}$ at infall at redshift $z=0$ are considered and weighted with their initial stellar mass $m_\mathrm{star,\,init}$. We will use the virial radius at infall as a scale for the dark matter halo during the satellite evolution and we will refer to it just as the virial radius $r_{200}$. The star formation before infall has been further investigated in \citep{maccio2017}. The galaxies show various behaviours of star formation in isolation, from absence of star formation in \textit{satV} to a starburst due to a merger with substructure in \textit{satII}. On the orbit however the star formation plays no or at least a minor role.

\subsubsection*{Metallicity}
\label{sec:metals}

%%%%%%%%%%%%%%%%%%%%%%%%%%%%%%%%%%%%%%%%%%%%%%%
\begin{figure}
\includegraphics[width=0.47\textwidth]{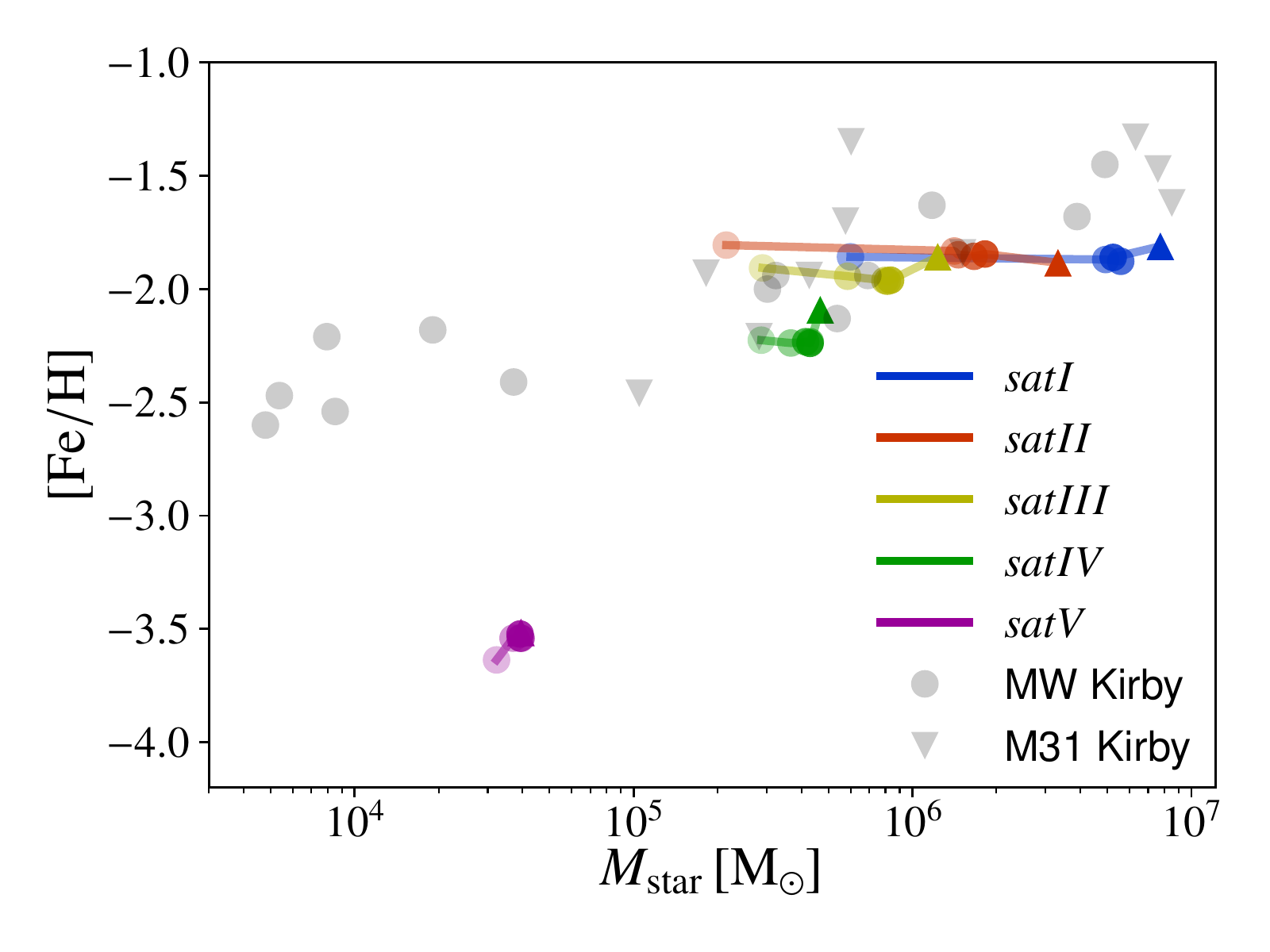}
\vspace{-.35cm}
\caption{Metallicity as a function of the stellar mass. The coloured triangles denote the isolated simulations while the filled circles denote the different orbits. The more violent the orbit, the fainter is the colour of the dots. Observations of Milky Way and M31 satellites from \citep{kirby2014} are shown as grey dots and triangles, respectively. }
\label{fig:metals}
\end{figure}
%%%%%%%%%%%%%%%%%%%%%%%%%%%%%%%%%%%%%%%%%%%%%%%

We present the mass weighted stellar metallicity as a function of stellar mass enclosed in a sphere of three 2D half mass radii in Fig. \ref{fig:metals}. 
Where the 2D half mass radius is defined as the radius of a cylinder along the $z$-axis (our line-of-sight) containing half of the galaxy stellar mass.

As before the triangle marks the position of the isolated simulation, while the different coloured circles mark the results for the different orbits, with faint colours being associated with more
disruptive orbits.  The grey circles and triangles represent observational results for the Milky Way and the M31 galaxy respectively. 
Our isolated runs nicely reproduce the observational trend from \citet{kirby2014}, with the exception of \textit{satV}, the apparent failure of this satellite is related 
to the very short time scale of star formation compared with the time-step of the simulation, which does not allow for a proper treatment of the enrichment (see PaperI for a thorough explanation of this issue). 

When the satellites are exposed to the presence of a central halo, they do lose stellar mass (as expected), but they still move parallel to the relation, with an almost constant metallicity, due to the very low metal gradient in their stellar population.

%The grey dots and triangles show observational data by \citet{kirby2014} of Milky Way and M31 satellites, respectively. The colour scheme for the simulational data of this work is as explained above (see discussion of Fig. \ref{fig:moster}). The data from \citet{kirby2014} describe a more or less tight realtion in the metallicty-stellar mass plane because the enrichment is related to the amount stellar lifecycles. With the exception of an outlier, \textit{satV}, which has a much too low metallicity compared to data in its stellar mass range the simulated satellites lie largly in the band given by the observations. We think that the reason for the deviation of \textit{satV}, whose stellar mass is produce during a single burst, from the relation is the insufficient modelling of stellar enrichment during single stellar bursts in our feedback and star formation recipe. This matter is treated more in detail in \citet{maccio2017}. Since there is not much of a metallicity gradient in the dwarf galaxies from our sample and the stellar populations are well mixed, we dont see much of a change in metallicity due to stripping. The loss of stellar mass lets the satellites evolve in the different orbits more or less along the scaling relation of the data. That may be a reason for the little scatter along the relation in contrast to Fig. \ref{fig:sig_r12}. 

\subsubsection*{Stellar kinematics}
\label{sec:stellar_kin}

%%%%%%%%%%%%%%%%%%%%%%%%%%%%%%%%%%%%%%%%%%%%%%%
\begin{figure}
\includegraphics[width=0.47\textwidth]{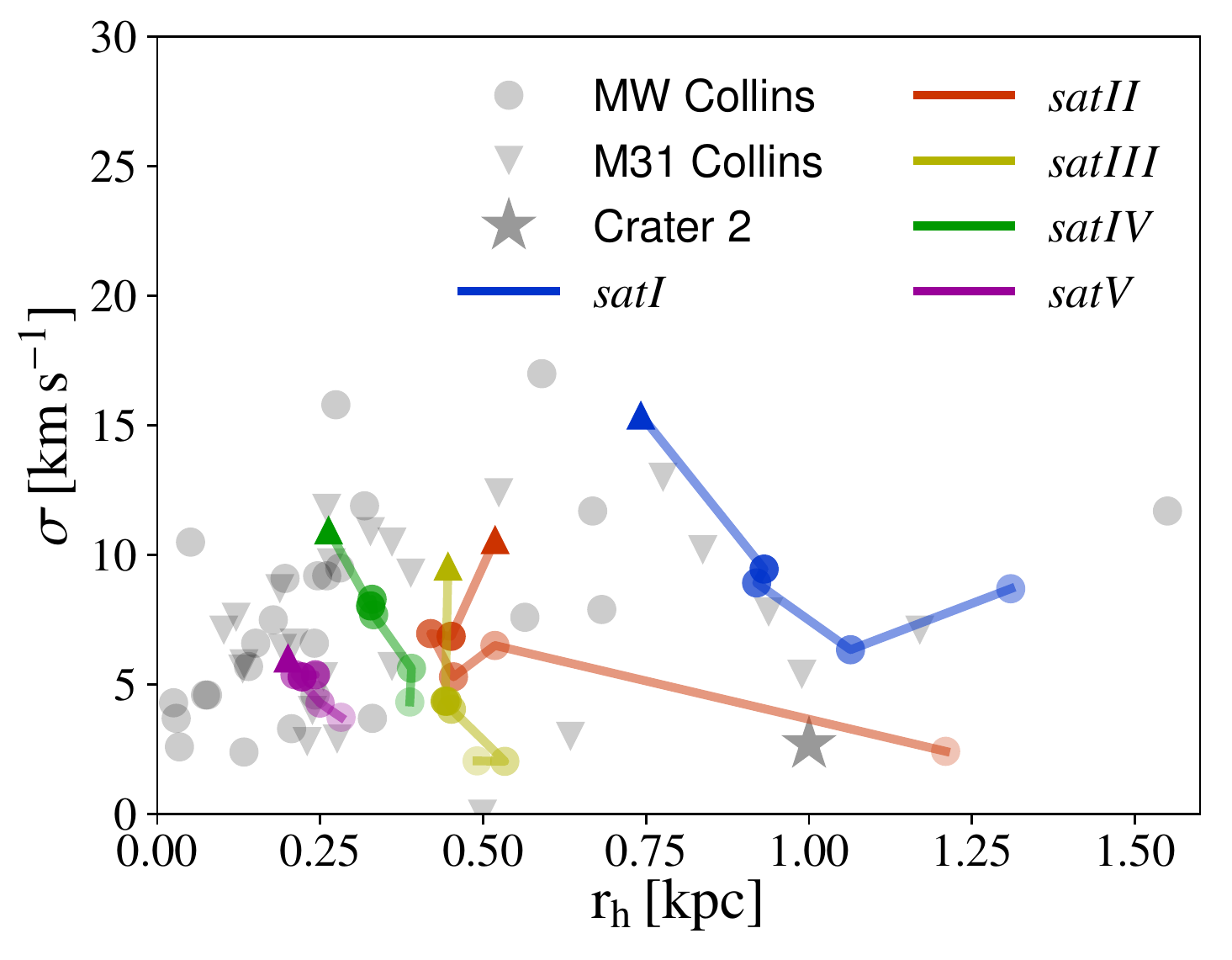}
\vspace{-.35cm}
\caption{Projected velocity dispersion inside the stellar 2D half mass radius as a function of the stellar 2D half mass radius. coloured triangles denote the isolated simulations while the filled circles denote the different orbits. The more violent the orbit, the fainter is the colour of the dots. Observations of Milky Way and M31 satellites (for references see section \ref{sec:stellar_kin})
 are shown as grey dots (and a star for the recently discovered satelite Crater 2) and triangles, respectively. }
\label{fig:sig_r12}
\end{figure}
%%%%%%%%%%%%%%%%%%%%%%%%%%%%%%%%%%%%%%%%%%%%%%%

%%%%%%%%%%%%%%%%%%%%%%%%%%%%%%%%%%%%%%%%%%%%%%%
\begin{figure*}
\includegraphics[width=\textwidth]{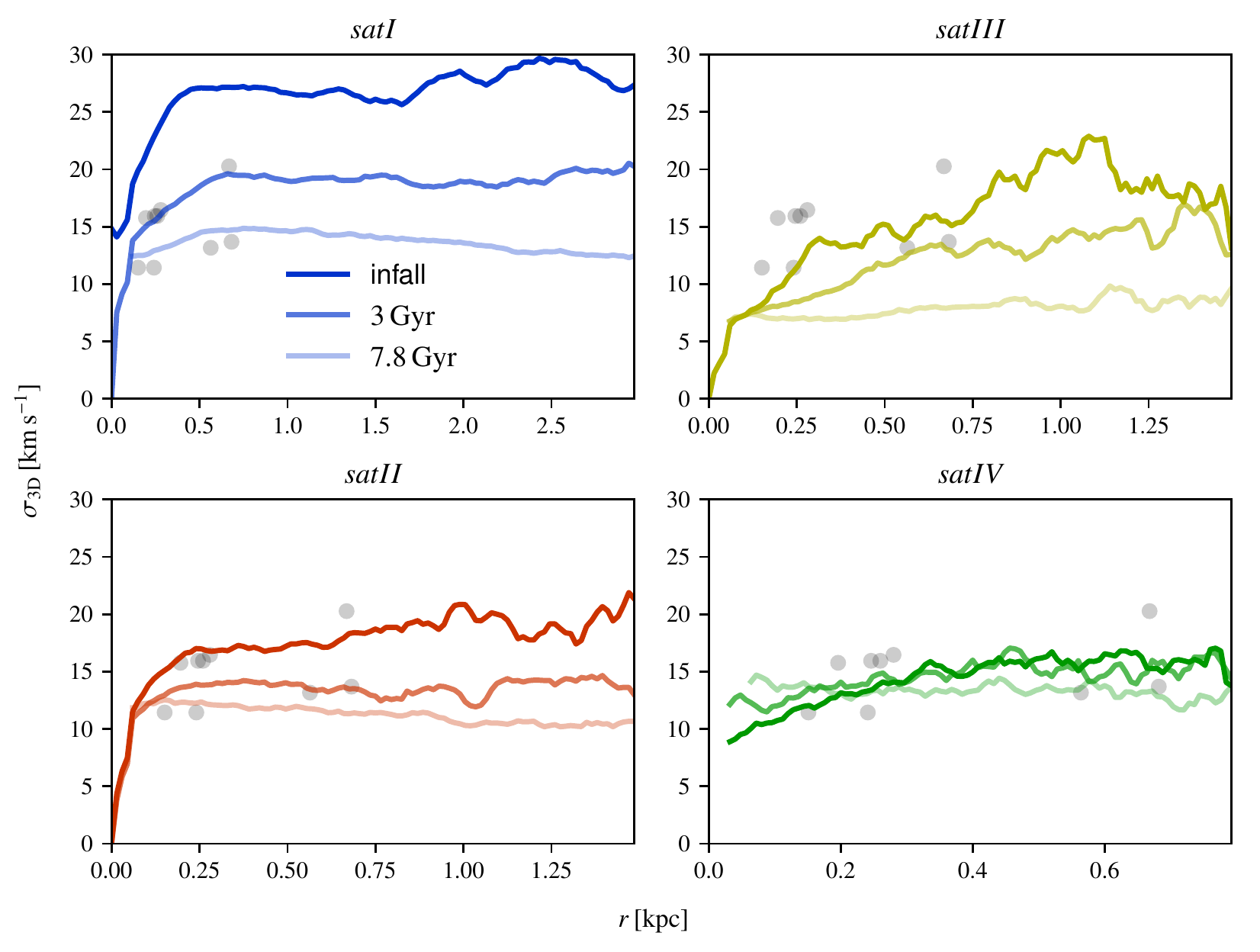}
\vspace{-.35cm}
\caption{Three dimensional velocity dispersion of the stellar particles on \textit{orbitII} evaulated at infall (solid), $3\Gyr$ and $7.8\Gyr$ ($z\approx0$) (faint, solid). Grey points denote line of sight velocity dispersion measurements from \citet{Walker2009} of the nine most massive Milky Way satellites rescaled by a factor of $\sqrt{3}$ \citep{wolf2010}.}
\label{fig:veldispprof}
\end{figure*}
%%%%%%%%%%%%%%%%%%%%%%%%%%%%%%%%%%%%%%%%%%%%%%%

The effect of accretion onto a more massive galaxy is instead clearly visible in the velocity dispersion-size relation which is shown in Fig. \ref{fig:sig_r12}.
The size ($r_{\rm h}$) is again the 2D half mass radius already introduced above, while the velocity dispersion is the 1D dispersion along the line of sight (the $z$=axis in our case) computed
within $r_{\rm h}$.
In this plot the observational data are represented by grey dots and triangles for Milky way and M31 satellites, respectively. They are taken from a compilation from M.Collins (private communication) including data from \citet{Walker2009} for the Milky Way and \citet{Tollerud2012,Tollerud2013,Ho2012, Collins2013, Martin2014} for M31 satellites. The size and dispersion measurements of the recently discovered satellite Crater 2 are taken from \citet{Caldwell2017}.

Our isolated haloes (triangles) lie well within the relation as were our initial conditions (see Paper I). Stripping and tidal forces modify both the size and the velocity dispersion of
the galaxies, which, depending on the orbit,  at redshift zero tend to occupy the whole space covered by the observations.

It is interesting to note that simulated galaxies with larger sizes tend
to have a larger deviation (both in size and dispersion) from the isolated
runs, suggesting that small galaxies tend to be more resilient to tidal effects.
Overall the scatter in our simulated size-dispersion velocity is
in very good agreement with the observed one. 
{Further we want to emphasise that \textit{satII} and \textit{orbitV} end up with an extremely low velocity dispersion at a half mass radius of about $1.2\kpc$. This is in very good agreement with the properties of the recently discovered Crater2 satellite \citep{Caldwell2017}.
This implies that the formation of such extended and cold structures is not a challenge for the current LCDM model, which can be explained 
as highly perturbed objects (see also \citet{Munshi2017}).

%This is in very good agreement with the observed scatter in the size-$\sigma$ relation which does increase with galaxy size. 

To better understand the time evolution of the stellar kinematics  in our satellites, in Fig.  \ref{fig:veldispprof} we show the radial profile of the three dimensional stellar velocity dispersion on \textit{orbitII} at three different 
times: at infall and after $3$ and $7.8\Gyr$ (corresponding to redshift $z=0$). Here we only consider \textit{satI} to \textit{satIV} because \textit{satV} has not sufficient stellar particles to resolve the kinematics properly (see Table \ref{tab:sample}).
In the same plot we also show as reference (grey circles) the line of sight velocity dispersion measurements for the nine most massive Milky Way satellites at the half light radius from \citet{Walker2009} 
rescaled by a factor of $\sqrt{3}$ \citep[see][]{wolf2010}).
As time goes by the mass (DM+stellar) loss causes an overall decrease of stellar velocity dispersion, which also tends to become more isothermal, with a very flat distribution
(see for example the case for \textit{satIII}) at redshift zero. We ascribe this effect to the  particle phase space mixing due to tidal effects, which seems to ``thermalize'' the galaxy.

%%%%%%%%%%%%%%%%%%%%%%%%%%%%%%%%%%%%%%%%%%%%%%%
\begin{figure*}
\includegraphics[width=\textwidth]{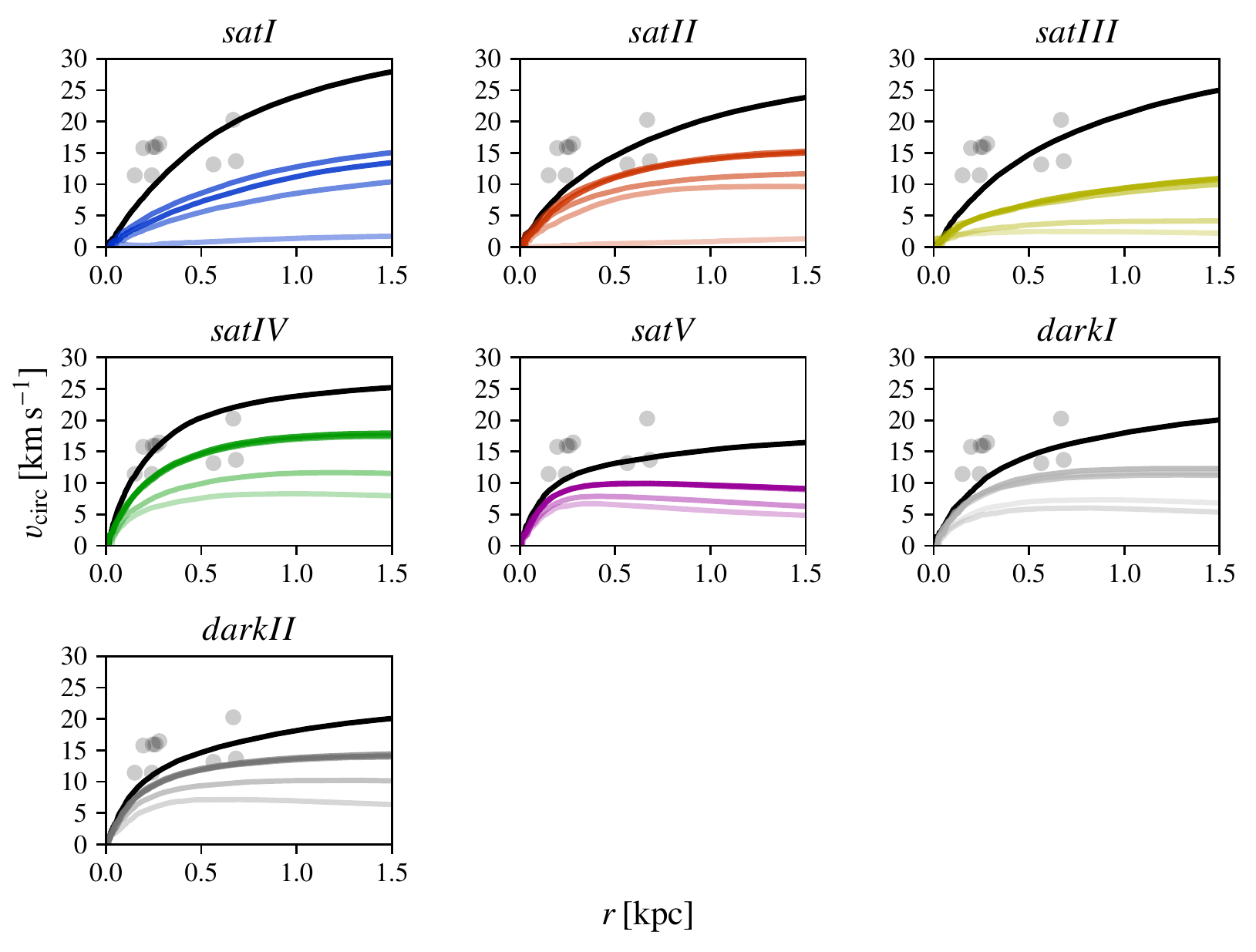}
\vspace{-.35cm}
\caption{Circular velocity profiles for the five luminous satellites for the isolated run (black) and the orbits (colour coding as before). The more violent the orbit, the fainter is the line colour. Grey points denote line of sight velocity dispersion measurements from \citet{Walker2009} of the nine most massive Milky Way satellites rescaled by a factor of $\sqrt{3}$ \citep{wolf2010}.}
\label{fig:circ1}
\end{figure*}
%%%%%%%%%%%%%%%%%%%%%%%%%%%%%%%%%%%%%%%%%%%%%%%

Another interesting (and measurable) quantity to look at is the circular velocity, defined as $v_\mathrm{circ}=\sqrt{\frac{G M(<r)}{r}}$, where $M(<r)$ is the total mass enclosed in a sphere with radius $r$ around the centre. In Fig. \ref{fig:circ1} we show the final ($z=0$) circular velocity radial profile for all our seven satellites and for different orbits. 
In each panel the black line represents the isolated run, while the coloured lines are the orbit runs, with, as before, fainter colours for more disruptive orbits.
We also show, as in the previous figures, the observations of the nine most massive Milky Way satellites  as an orientation \citep[data from][]{Walker2009}.
As already noted in previous studies, circular velocity profiles can be significantly lowered in the inner few hundred parsecs, 
even without losing a large amount of mass on these scales (see Fig. \ref{fig:moster}). The profiles also tend to evolve in a sort of 
self-similar way, preserving their initial shape, with the exception of the most extreme orbit.

\label{sec:stellar_kin}

\subsection{Evolution of the dark matter profile}
\label{sec:dm_profile}

%%%%%%%%%%%%%%%%%%%%%%%%%%%%%%%%%%%%%%%%%%%%%%%
\begin{figure*}
\includegraphics[width=\textwidth]{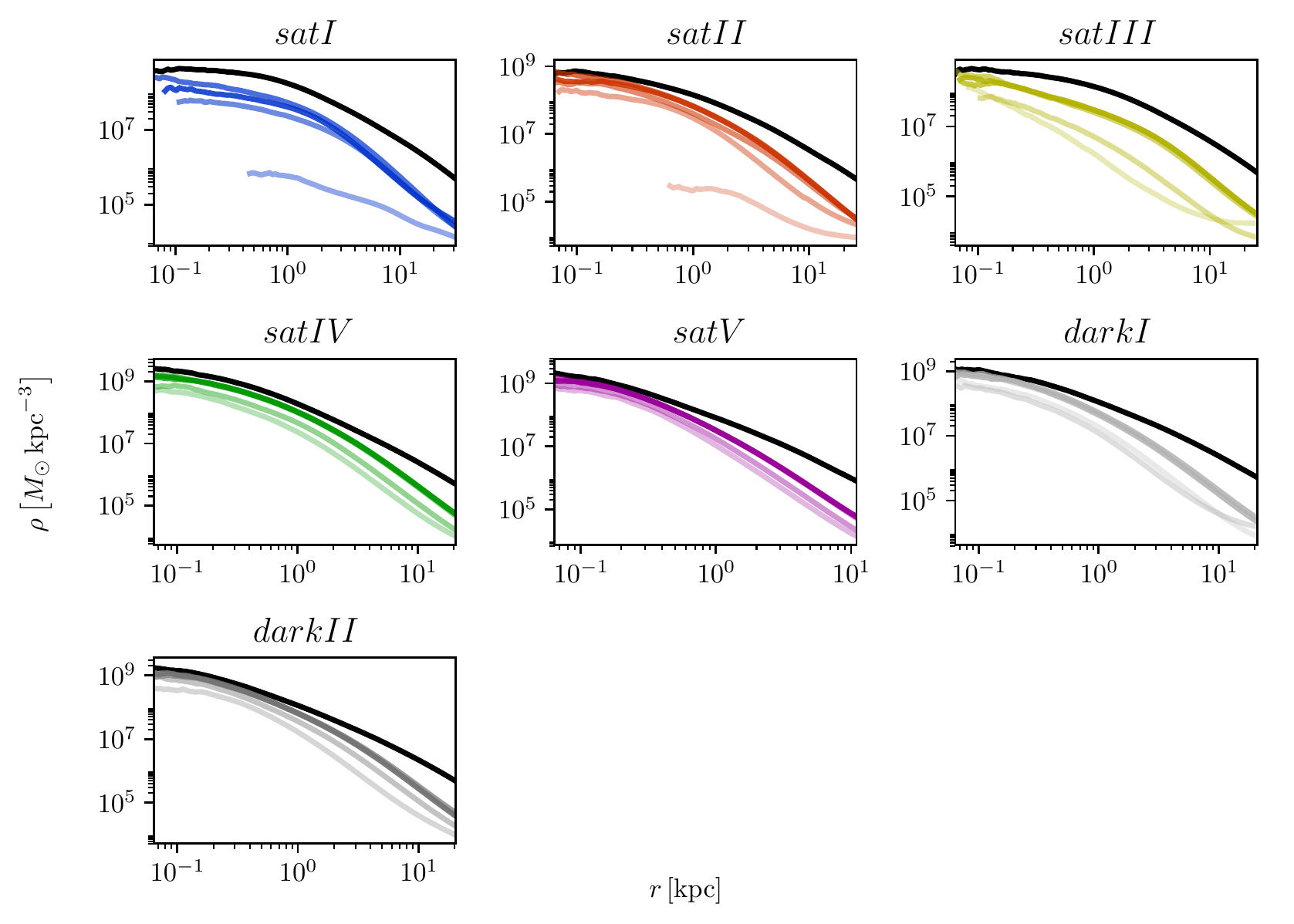}
\vspace{-.35cm}
\caption{Dark matter density profile of all the satellites in isolation (black) and on the orbits (colour coding as before) at redshift $z=0$. The more violent the orbit, the fainter is the line colour. The profiles are shown from twice the softening length up to the virial radius at infall.}
\label{fig:densprofs}
\end{figure*}
%%%%%%%%%%%%%%%%%%%%%%%%%%%%%%%%%%%%%%%%%%%%%%%

We now turn our attention to the dark (matter) component of our satellites. 
In Fig.  \ref{fig:densprofs}  we show the redshift zero dark matter distribution for all our galaxies in the various runs at redshift $z=0$, as before the black line represents the isolation run,
while the coloured lines are for the different orbits. The profiles are shown from twice the gravitational dark matter softening to the virial radius at infall (the exceptions are satellites \textit{satI}, \textit{satII} and \textit{satIII} on the most extreme orbit, since they end up  with a very  low dark matter content at $z=0$ and which pushes
the convergence radius to larger scales).
As expected the tidal stripping due to the central potential is stronger in the outer parts of the profile, which depart more from the isolation case.
On the other hand the stripping does not happen in an ``onion-like'' fashion, with the outer part being progressively removed while the centre remains unaltered.
On the contrary, the whole profile reacts to the stripping and the central density is also lowered even on the most mild orbits. 
This global reaction can be ascribed to the typical box orbits of dark matter particles \citep[e.g.][]{Bryan2012}, which allow particles in the centre
at a given time step, to spend quite some time in the outskirts of the halo at a subsequent time, and hence are prone to be stripped.

As already described in PaperI, some of our satellites (\textit{satI, satII, satIII} as shown by the isolation runs) 
start with a cored dark matter profile (or at least a profile with a shallower slope than NFW), this is due to the fact that for those satellites, the star formation rate is vigorous  enough to create large gas outflows, which in turn flatten the dark matter profile \citep{Read2005, Pontzen2012, Maccio2012, DiCintio2014a, Onorbe2015}. 
Fig. \ref{fig:densprofs}  seems to suggest a steepening of the profile during its evolution. To better look into this possibility we plot in Fig. \ref{fig:alphas} the inner logarithmic dark matter density slope $\alpha$ as a function of stellar mass. The slope $\alpha$ is computed between 1-2\% of the initial (at infall) virial radius, following
\cite{Tollet2016} and PaperI; the triangles mark the isolation runs, while the circles represent the different orbits, finally the stellar mass for the two dark haloes has been arbitrarily set. 

No matter if the halo contained stars or not, or whether it starts with a flattened profile (blue and yellow symbols) or with a cuspy one (green symbols), in all cases
the effect of accretion is to steepen the dark matter profile. It is important to notice that the profile steepening is not 
due to a contraction of the halo but it is due to a slightly stronger mass removal in the outer regions of the halo with respect to 
its very centre.

In Fig. \ref{fig:densprofs} we do not show the final density profile slope for  \textit{satI}  (blue), \textit{satII} (red) and \textit{satIII}  (yellow) for the most 
 extreme orbit. This is because at $z=0$ these satellites do not have a clear dark matter centre to build the profile. 
 It is nevertheless interesting to look at the evolution of the dark matter profile as a function of time before the satellite disruption.
This is shown in Fig. \ref{fig:alphas_time} where we present the difference of the density slope $\alpha$ w.r.t. to the isolation case
as a function of time: the plots show \textit{satI} (blue line), \textit{satII} (red line) and \textit{satIII} (yellow line) on \textit{orbitIV} (for \textit{satI})  and \textit{orbitV} (for \textit{satII} and \textit{satIII}), respectively, 
and the profile slopes are  averaged over five time-steps to reduce the noise. 
It is evident that the steepening of the profile is  present even for satellites that are completely shredded apart by tidal forces.

% We do not show the data points for \textit{satI}, \textit{satII} and \textit{satIII} on their most extreme orbits in Fig. \ref{fig:alphas} because the dark matter in \textit{satI} and \textit{satII} is already so diluted in the centre that the profile is essentially flat ($\alpha=0$) and \textit{satIII} developes a slope of $\alpha=-2.48$ (see also Fig. \ref{fig:densprofs})	which is far out of the range of slopes we want to show here. To show that the satellites still develop a steep central dark matter slope before they are totally stripped of dark matter, we present the time evolution of $\alpha$ of \textit{satI} (blue line) and \textit{satII} (red line) on \textit{orbitIV} and \textit{orbitV}, respectivley, in Fig. \ref{fig:alphas_time}. On the $y$-axis we show the deviation from the central slope in the isolated run $\alpha_\mathrm{iso}$. The data is averaged over five timesteps to reduce the scatter. The star at the end of each track marks the time where there are less than $150$ dark matter particles left in a sphere of a radius of $1\%\,r_{200}$. If the data are below the black horizontal line indicating $\alpha=\alpha_\mathrm{iso}$, the profile of the satellite is steeper than isolated run and flatter above the line. %It seems like on the violent orbits either, if there is a core in the dark matter profile, the centre becomes totally diluted of dark matter or, if there is no core, the dark matter developes an incredibly steep central slope.

When the results of PaperI and this work are combined, they imply that the observational discovery of a dark matter core in one of the low mass satellites of our
own Galaxy will strongly challenge the predictions of the \LCDM model. It will be very difficult to explain such dark matter core invoking the effect of baryons 
(Paper I and triangles in Fig. \ref{fig:alphas}) or the effect of environment and accretion (again Fig. \ref{fig:alphas}).
The discovery of a flat dark matter distribution will then be an indication of a different nature for dark matter: warm \citep[but see][]{Maccio2012b}, self interacting 
\citep{Vogelsberger2014b,Elbert2015} or even more exotic models \citep[e.g.][]{Maccio2015}.

%%%%%%%%%%%%%%%%%%%%%%%%%%%%%%%%%%%%%%%%%%%%%%%
\begin{figure*}
\includegraphics[width=\textwidth]{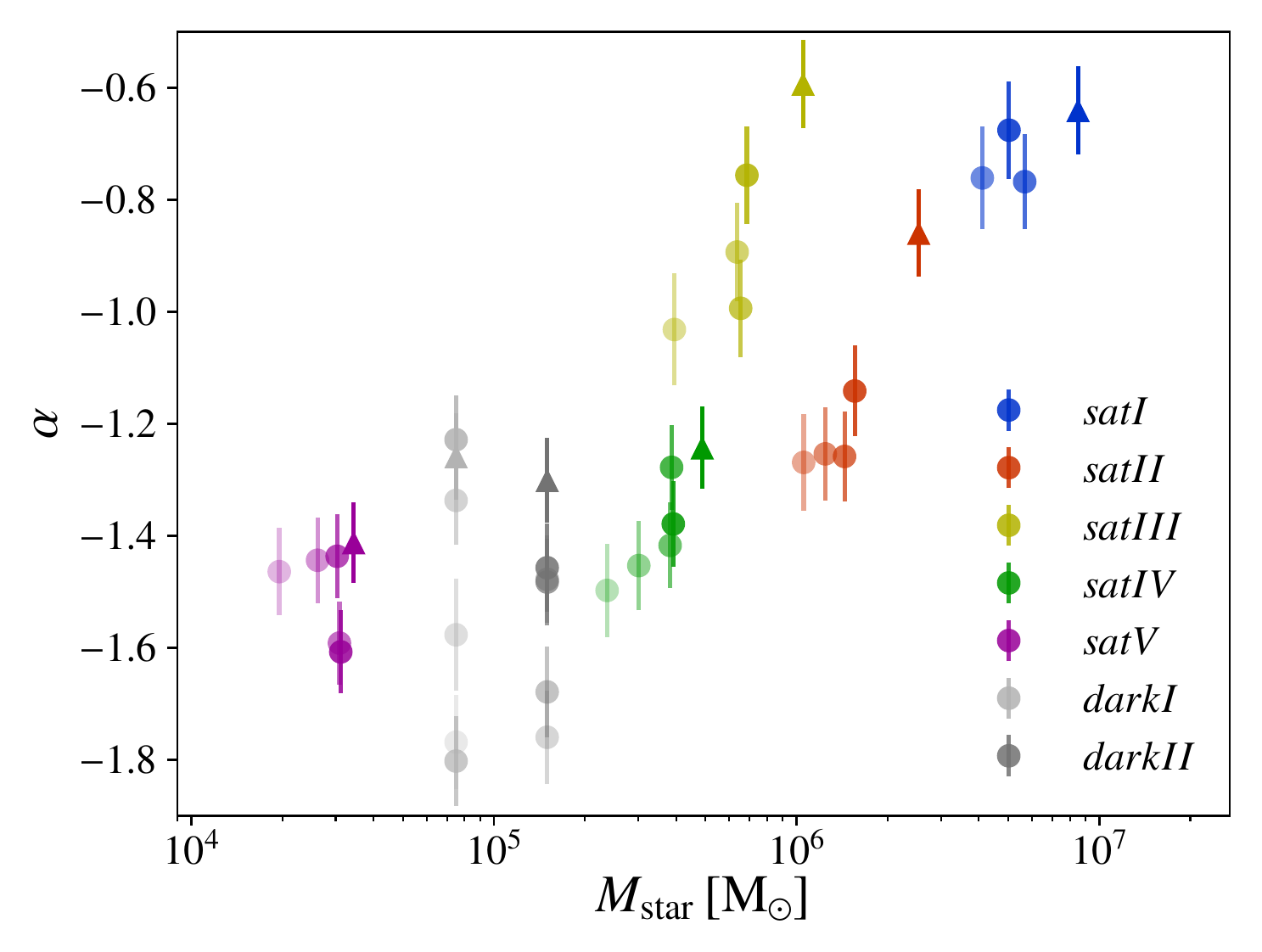}
\vspace{-.35cm}
\caption{Inner logarithmic slope of the dark matter density profile $\alpha$ fitted between $1\%$ and $2\%$ of the virial radius as a function of stellar mass inside three 3D half mass radii. Stars denote measurements at infall, while triangles denote the isolated simulations and filled circles the orbit runs at redshift $z=0$. The more violent the orbit, the fainter is the colour.}
\label{fig:alphas}
\end{figure*}
%%%%%%%%%%%%%%%%%%%%%%%%%%%%%%%%%%%%%%%%%%%%%%%
%%%%%%%%%%%%%%%%%%%%%%%%%%%%%%%%%%%%%%%%%%%%%%%
\begin{figure}
\includegraphics[width=0.47\textwidth]{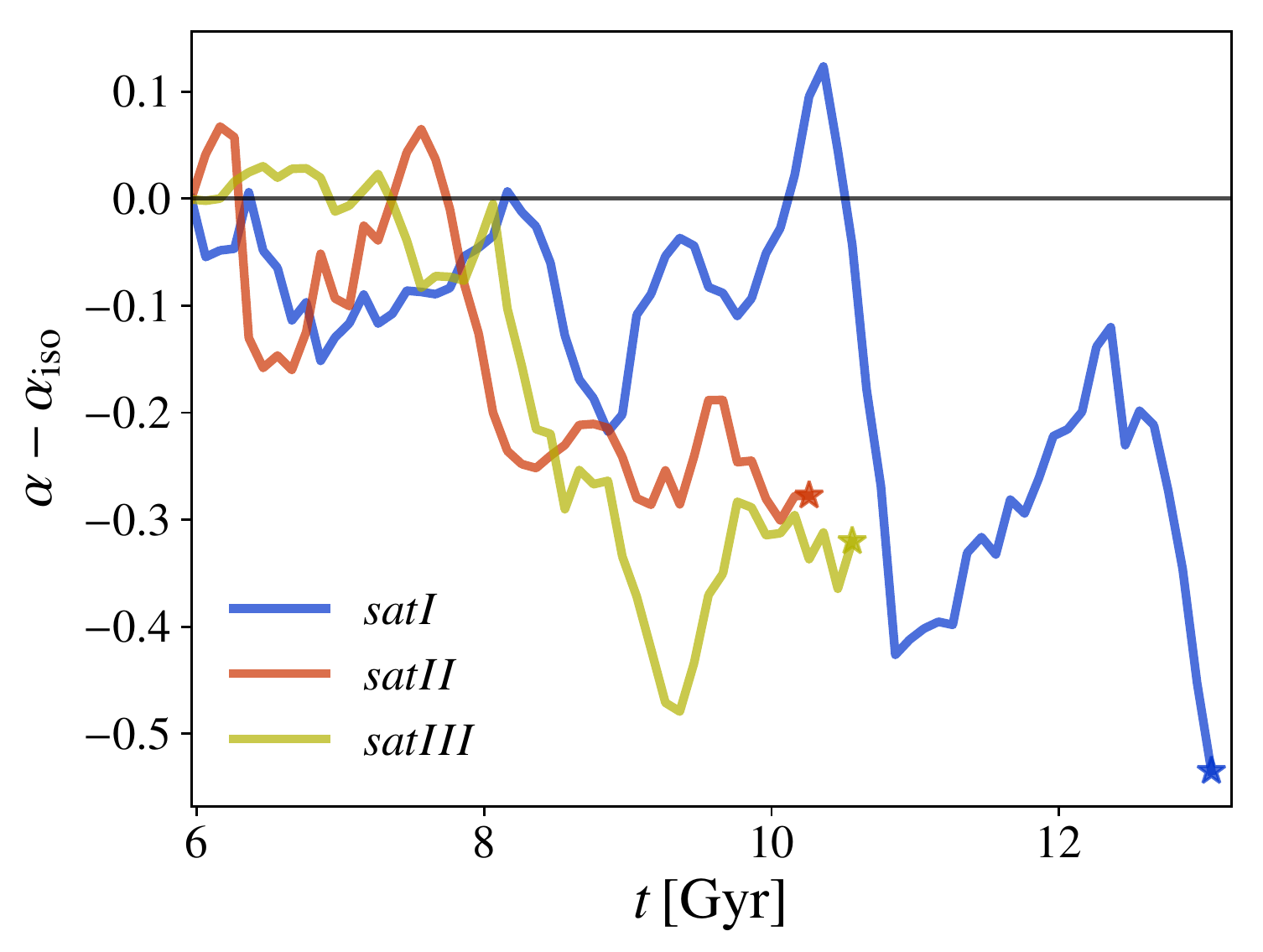}
\vspace{-.35cm}
\caption{Time evolution of the deviation of $\alpha$ on \textit{orbitIV} and \textit{orbitV} from $\alpha_\mathrm{iso}$ (for the isolated runs) for \textit{satI}, \textit{satII} and \textit{satIII}, respectively. The stars denote the time where there are less than $150$ dark matter particles left inside a sphere of a radius of $1\%\, r_{200}$.}
\label{fig:alphas_time}
\end{figure}
%%%%%%%%%%%%%%%%%%%%%%%%%%%%%%%%%%%%%%%%%%

\subsection{Central dark matter density slope and satellite survival}

Previous works studying environmental effects on satellite galaxies \citep[e.g.][and references therein]{Kazantzidis2004b, Pennarubia2010} have shown
that satellites with cored dark matter density profiles are more easily stripped and disrupted than cuspy ones.
The results shown in Fig. \ref{fig:alphas} seemed to confirm such a correlation also in our cosmologically based simulations, but we want to be more quantitative. 

In Fig.  \ref{fig:mfrac_alpha_dm} we show the ratio of the dark matter enclosed within three half mass radii for the different satellites on the different orbits
with respect to the isolated case as a function of the initial (at infall) density profile slope at redshift $z=0$. 
It is quite evident that on every orbit, cuspy satellites like \textit{satV} ($\alpha_{\rm infall} = -1.5$) are able to retain a larger fraction of their initial mass than 
cored ($\alpha_{\rm infall} > -1.0$ ) satellites like \textit{satI} and \textit{satIII}. These cored satellites lose more than 70\% of their initial (dark) mass even on the more gentle orbit (darker colours)
and up to almost 100\% on the most extreme ones (fainter colours).

%this correlation in our set of simulations we have looked at the mass loss as a function of the central dark matter density slope at infall $\alpha_\mathrm{infall}$ in Fig. \ref{fig:mfrac_alpha_dm} and Fig. \ref{fig:mfrac_alpha_all}. In the first figure we show the dark matter mass enclosed in a constant sphere with a radius of three half mass radii at infall in terms of mass in isolation on the $y$ axis. The data for all five orbits is shown with decreasing opacity for the more disruptive orbits. We can clearly confirm the correlation as on all orbits there is a strong trend of enhanced stripping for the satellite with the flatter profiles.

The different initial dark matter density slope also affects the stellar mass loss, since the dark matter acts as shield for the stars.
In Fig. \ref{fig:mfrac_alpha_all} we plot the mass loss as defined in Fig. \ref{fig:mfrac_alpha_dm} as function of the initial dark matter profile density slope
but this time for stars and dark matter. In order to avoid a too crowded plot we only show results for \textit{orbitIV}, which is the most disruptive orbit for which 
all five satellites still have a well defined centre. 

In general stars are more resilient to tidal forces than dark matter, and this is due to their smaller spatial extent and larger stellar density;
for example  \textit{satIII}  is able to retain 40$\%$ of its stars while it is practically totally stripped of dark matter.
Nevertheless there is still a correlation between the stellar mass loss and the initial dark matter slope. An exception to this relation seems to be given by \textit{satI}, which has a very strong mass loss
both in the stellar and dark matter components despite having a similar $\alpha_{\rm infall}$ as \textit{satIII}. 
This is due to the different slope for the {\it stellar} density profiles between \textit{satI} and \textit{satIII}, while the first satellite has a slope of 
$-1.80$ (again evaluated between 1 and $2\%$ of the virial radius),  the stellar slope for \textit{satIII} (and all the other satellites) it is  close to $-3.0$. 
This difference in stellar slope is most likely related to a major merger event that occurred for \textit{satI} shortly before redshift one and strongly reshuffled the stellar
particle orbits. 

Finally we note that some satellites (especially \textit{satI}) are left at $z=0$ with practically no dark matter in their central region, where they are fully stellar dominated, 
resembling more an extended globular cluster than a dwarf galaxy, we plan to look more into this issue in a forthcoming paper.

\begin{figure}
\includegraphics[width=0.47\textwidth]{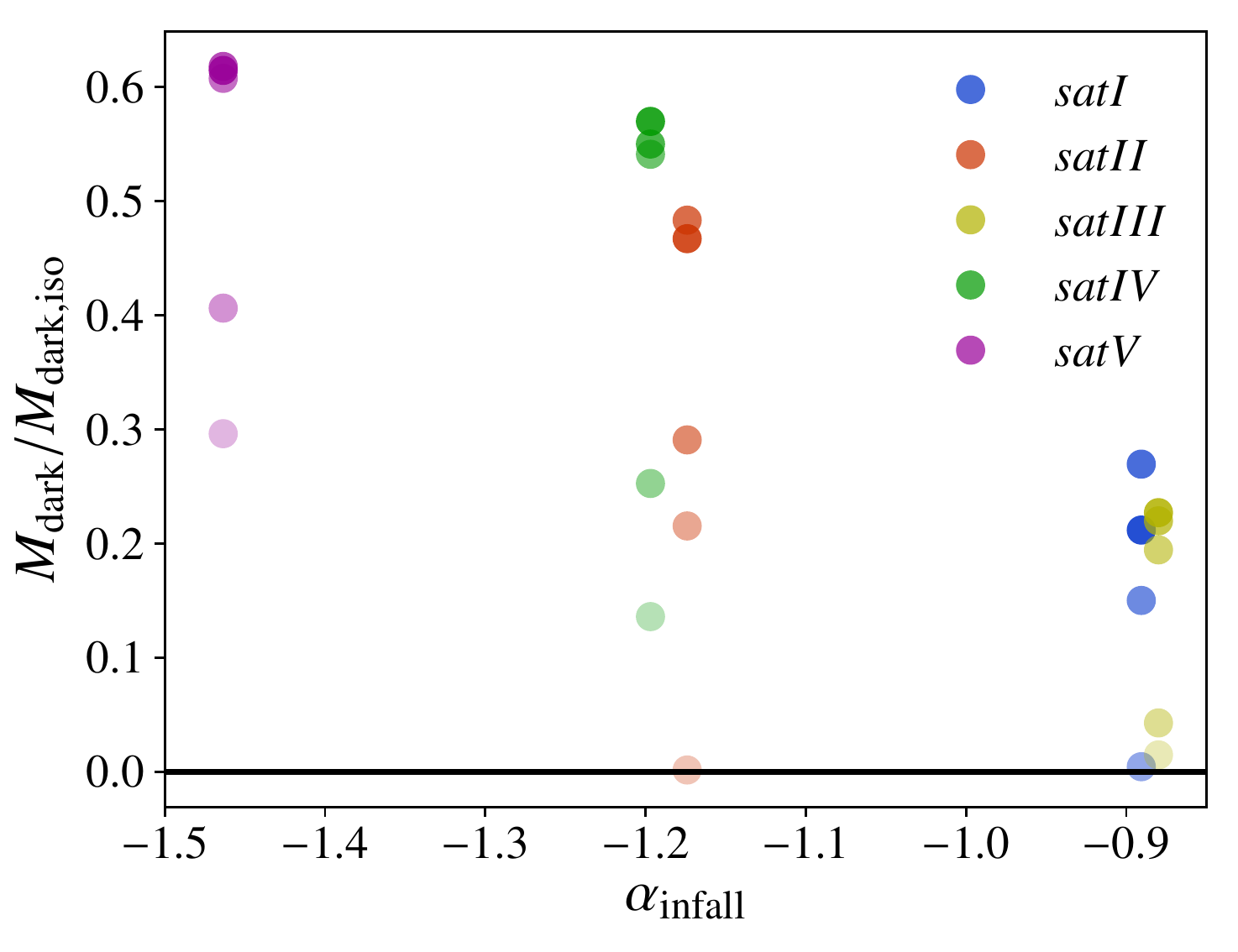}
\vspace{-.35cm}
\caption{Dark matter mass enclosed in a constant sphere with a radius of three stellar half mass radii at infall on the individual orbits compared to the isolated run as a function of the central dark matter density slope at infall $\alpha_\mathrm{infall}$.}
\label{fig:mfrac_alpha_dm}
\end{figure}
%%%%%%%%%%%%%%%%%%%%%%%%%%%%%%%%%%%%%%%%%%
%%%%%%%%%%%%%%%%%%%%%%%%%%%%%%%%%%%%%%%%%%%%%%%
\begin{figure}
\includegraphics[width=0.47\textwidth]{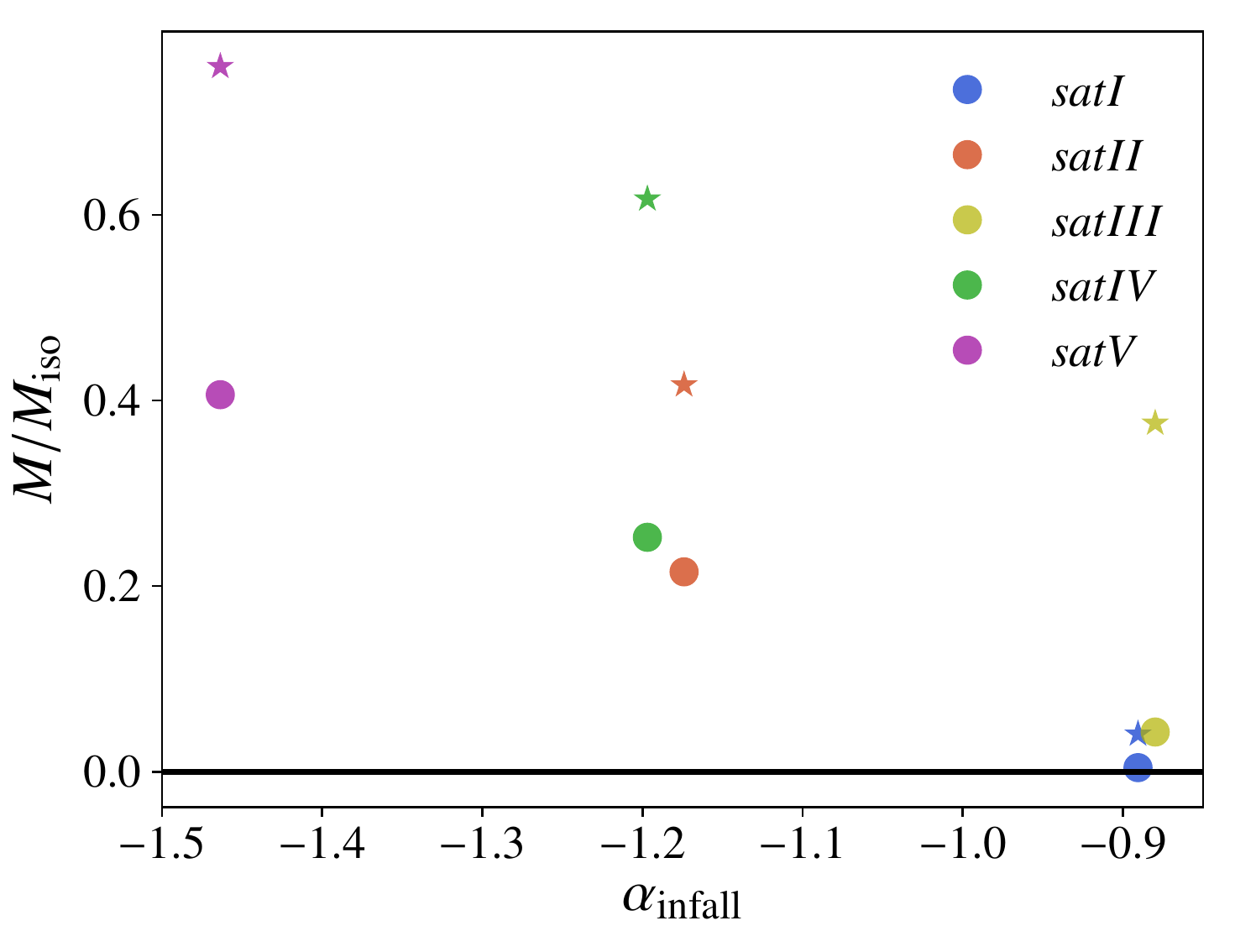}
\vspace{-.35cm}
\caption{Dark matter (dots) and stellar (stars) mass enclosed in a constant sphere with a radius of three stellar half mass radii at infall on \textit{orbitIV} compared to the isolated run as a function of the central dark matter density slope at infall $\alpha_\mathrm{infall}$.}
\label{fig:mfrac_alpha_all}
\end{figure}
%%%%%%%%%%%%%%%%%%%%%%%%%%%%%%%%%%%%%%%%%%

\section{Discussion and Conclusions}
\label{sec:discussion} 

This work is the second of a series of papers in which we are trying to understand the formation and evolution of the smallest galaxies in the universe.

In the first paper \citep{maccio2017} we presented a series of 27 cosmological simulations of halos in the mass range $5\times 10^8$ to $10^{10}\,$ \Msun, these simulations were run
till $z=1$ and aimed to describe the properties of satellite galaxies before accretion.

In this second paper we used a subsample of these cosmological simulations as initial conditions for a series of binary  hydrodynamical simulations (satellite + host) 
with the goal of understanding the effects of accretion and environment on satellite galaxies.

More specifically we used a total of seven haloes (5 luminous and 2 dark) with a virial mass in the range $4\times 10^9$ to $10^{10} \Msun$. We modeled
the central halo with an analytic potential of halo (NFW) plus disc (Miyamoto \& Nagai), and we run the simulations from redshift one until redshift zero.
We use five different orbits (plus a radial orbit) for a total of 42 simulations. 
Each galaxy is also evolved ``in isolation'' meaning without the presence of the central halo for the same amount of time as our ``orbit'' runs.
We also add an analytic model for ram pressure and we test the results
of our analytic potential against a live halo.

We find that our cosmological initial conditions differ from model (pre cooked) galaxies in their kinematics: our galaxies, even before accretion do not have
a well defined rotating stellar disc and are dispersion supported, with an average value of $v_{\varphi} / \sigma$ of 0.14, and as low as 0.04.

While orbiting around the central host, all galaxies lose a considerable fraction of their halo mass, as a consequence they drift away from the widely used
abundance matching relations, due to a reshuffling of the mass rank order of the satellites.
Only the more extreme orbits, with small pericentre distances are effective in stripping the stellar component too, 
and the stripping is substantial only for our most massive (and extended) galaxies.
Star formation is strongly suppressed after the infall (in comparison with the isolation run), this happens regardless of the presence or absence
of ram pressure. 

The environment has different effects on different scaling relations. In the stellar mass - metallicity plane, galaxies, even when they lose stellar
mass, tend to keep an almost constant metallicity, due to the lack of strong metallicity gradients in the galaxy.
In the size - velocity dispersion relation, galaxies move substantially in all directions, depending on their orbits and initial size.
Galaxies tend to become more extended and to reduce their velocity dispersion (due to dark matter stripping).
This explains the overall large observed scatter in the $r-\sigma$ plane.% and the increase of scatter with galaxy size.

Finally the interaction with the host leads to a flattening of the stellar velocity dispersion profiles of the satellites, possibly due to 
the potential perturbation redistributing the stellar orbits into a more ``thermalized'' system.

The dark matter component is also very strongly perturbed, first of all the removal of particles from the outer part of the halos does not
happen in an ``onion-like'' fashion, with the external layers removed first. The whole density profile reacts to the accretion and even 
the most internal regions are affected by mass removal even if at a lower degree of the external ones.
This implies an overall suppression of the circular velocity curves with respect to the isolation runs, even in the central part.

We find a correlation between the initial (at infall) inner dark matter density slope and the efficiency of mass removal and satellite survival that is in good agreement with previous studies. Cored satellites are less resilient to stripping and tidal forces and are more
prone to lose a very large fraction of their dark matter mass (if not all of it) on orbits with close pericentre passages.
Some cored satellites are so heavily stripped in their dark matter component that they end up being almost entirely stellar dominated within
three stellar half mass radii. Stars are also more easily stripped when embedded in a halo with a flat density profile, even though to a lower extent. 

%and as a result the survival of the satellites dark matter halo. For the efficiency of stellar stripping however we find that the correlation is much more relaxed and even satellites with dark matter profiles that initially are fallter than NFW can leave a confined galaxy behind that has essentially no dark matter counterpart. This might provide another formation mechanism for globular clusters in the Milky Way.}

We witness a steepening of the central slope of the dark matter profile  during accretion, with more extreme orbits ending up 
with the most cuspy dark matter profiles. 
Even profiles that initially (before accretion) have a dark matter density profile shallower than NFW (due to baryonic effects) evolve into cuspy profiles with slopes consistent with pure N-body simulations when set on orbits with small pericenter passages (7-20 kpc).
Interestingly also our dark haloes (without stars) do steepen their profiles.

Overall our simulations seem to make a quite clear prediction of steep dark matter profiles for objects at the edge of galaxy formation, 
a prediction that, if falsified by observations, can force us to reconsider the collision-less and cold nature of dark matter.

\section*{Acknowledgements}
This research was carried out on the High Performance Computing resources at New York University Abu Dhabi; 
on the {\sc theo} cluster of the Max-Planck-Institut f\"ur Astronomie and on the {\sc hydra} clusters at the Rechenzentrum in Garching.
TB and AVM acknowledge funding from the Deutsche Forschungsgemeinschaft via the SFB 881 program 
``The Milky Way System'' (subproject A1 and A2). JF and AVM acknowledge funding and support through the graduate college {\em Astrophysics of cosmological probes of gravity} by Landesgraduiertenakademie Baden-W{\"u}rttemberg. JF and TB  are members of the International Max-Planck Research School in Heidelberg. AO acknowledges support from the German Science Foundation (DFG) grant 1507011 847150-0.
CP is supported by funding made available by ERC-StG/EDECS n. 279954
%%%%%%%%%%%%%%%%%%%%%%%%%%%%%%%%%%%%%%%%%%%%%%%%%%



\bibliographystyle{mnras}
\bibliography{ref} % if your bibtex file is called example.bib


%%%%%%%%%%%%%%%%%%%%%%%%%%%%%%%%%%%%%%%%%%%%%%%%%%

%%%%%%%%%%%%%%%%% APPENDICES %%%%%%%%%%%%%%%%%%%%%

%\appendix

%%%%%%%%%%%%%%%%%%%%%%%%%%%%%%%%%%%%%%%%%%%%%%%%%%

% Don't change these lines
\bsp	% typesetting comment
\label{lastpage}
\end{document}